\documentclass[manuscript]{acmart}
\usepackage{amsmath,amsfonts}
\usepackage{algorithmic}
\usepackage{algorithm}
\usepackage{array}
\usepackage{svg}
\usepackage[caption=false,font=normalsize,labelfont=sf,textfont=sf]{subfig}
\usepackage{textcomp}
\usepackage{stfloats}
\usepackage{url}
\usepackage{verbatim}
\usepackage{graphicx}
\usepackage{bm}
\usepackage{xcolor}
\usepackage{todonotes}
\usepackage{siunitx}

\AtBeginDocument{%
  }

\setcopyright{acmlicensed}
\copyrightyear{2025}
\acmYear{2025}
\acmDOI{XXXXXXX.XXXXXXX}

\acmJournal{TACO}
\acmVolume{0}
\acmNumber{0}
\acmArticle{0}
\acmMonth{0}

\begin{document}

\title{Characterization of Real Communication Patterns and Congestion Dynamics in HPC Interconnection Networks}

\author{Miguel Sánchez de La Rosa}
\affiliation{%
  \institution{Universidad de Castilla-La Mancha (UCLM)}
  \city{Albacete}
  \country{Spain}}
\email{miguel.sanchez@uclm.es}

\author{Gabriel Gomez-Lopez}
\affiliation{%
  \institution{Universidad de Castilla-La Mancha (UCLM)}
  \city{Albacete}
  \country{Spain}}
\email{gabriel.gomez@uclm.es}

\author{Alejandro Baviera}
\affiliation{%
  \institution{Universitat Politècnica de València (UPV)}
  \city{Valencia}
  \country{Spain}}
\email{abavgua@etsinf.upv.es}

\author{Jose Duro}
\affiliation{%
  \institution{Universitat Politècnica de València (UPV)}
  \city{Valencia}
  \country{Spain}}
\email{jodugo1@disca.upv.es}

\author{Francisco J. Andújar}
\affiliation{%
  \institution{Universidad de Valladolid (UVA)}
  \city{Valladolid}
  \country{Spain}}
\email{fandujarm@infor.uva.es}

\author{Jesus Escudero-Sahuquillo}
\affiliation{%
  \institution{Universidad de Castilla-La Mancha (UCLM)}
  \city{Albacete}
  \country{Spain}}
\email{jesus.escudero@uclm.es}

\author{Pedro J. Garcia}
\affiliation{%
  \institution{Universidad de Castilla-La Mancha (UCLM)}
  \city{Albacete}
  \country{Spain}}
\email{pedrojavier.garcia@uclm.es}

\author{Francisco J. Alfaro}
\affiliation{%
  \institution{Universidad de Castilla-La Mancha (UCLM)}
  \city{Albacete}
  \country{Spain}}
\email{fco.alfaro@uclm.es}

\author{Maria E. Gomez}
\affiliation{%
  \institution{Universitat Politècnica de València (UPV)}
  \city{Valencia}
  \country{Spain}}
\email{megomez@disca.upv.es}

\author{Julio Sahuquillo}
\affiliation{%
  \institution{Universitat Politècnica de València (UPV)}
  \city{Valencia}
  \country{Spain}}
\email{jsahuqui@disca.upv.es}

\author{José L. Sánchez}
\affiliation{%
  \institution{Universidad de Castilla-La Mancha (UCLM)}
  \city{Albacete}
  \country{Spain}}
\email{jose.sgarcia@uclm.es}

\author{Francisco J. Quiles}
\affiliation{%
  \institution{Universidad de Castilla-La Mancha (UCLM)}
  \city{Albacete}
  \country{Spain}}
\email{francisco.quiles@uclm.es}

\renewcommand{\shortauthors}{Sanchez de la Rosa et al.}

\begin{abstract}
The interconnection network is a key component of Supercomputers and Data centers, and its design must cope with the increasing communication demands of current applications and services; otherwise, it may become a system bottleneck. The most challenging network design issues are the topology, routing algorithm, flow control, and power efficiency. However, even the most efficient interconnection networks may suffer severe performance degradation due to congestion, especially under specific network traffic patterns generated by communication operations in high-performance computing~(HPC), deep learning training, or online data-intensive services. In this context, characterizing and modeling these communication operations and the network traffic patterns they generate is a fundamental challenge for studying their impact on network performance. This paper presents a methodology, based primarily on the VEF Traces framework, to characterize, model, and simulate the communication patterns of representative computing- and data-intensive applications. More precisely, we have extended the VEF traces framework with tools that enable us to characterize network congestion, either directly from VEF traces or via simulations. We have analyzed a set of VEF traces obtained from runs of NEST, GROMACS, LAMMPS, and PATMOS on several Supercomputers. In these studies, we identify potential congestion scenarios that arise in realistic network configurations when certain collective operations are performed.
\end{abstract}


\maketitle

\section{Introduction}
\label{sec:introduction}

Simulation allows network designers to model high-performance interconnection networks and evaluate their performance under workloads generated by the system's end nodes when running specific benchmarks and applications. 
It is therefore essential to accurately characterize the communication patterns of these workloads in real systems, so that realistic network traffic can be fed into network simulators. Although this approach is not new and some simulation frameworks were using it in the past (e.g., DUMPI traces in SST \cite{sst-dumpi} or VEF traces\cite{Andujar16jsc}), it has gained popularity in recent studies (e.g., Astra-sim \cite{won2023astrasim2} or ATHLAS \cite{shen2025atlahs}) because it enables network designers and researchers to identify bottlenecks and other undesired conditions before any real deployment.

\begin{figure}[!hb]
    \centering
    \includegraphics[width=0.85\columnwidth]{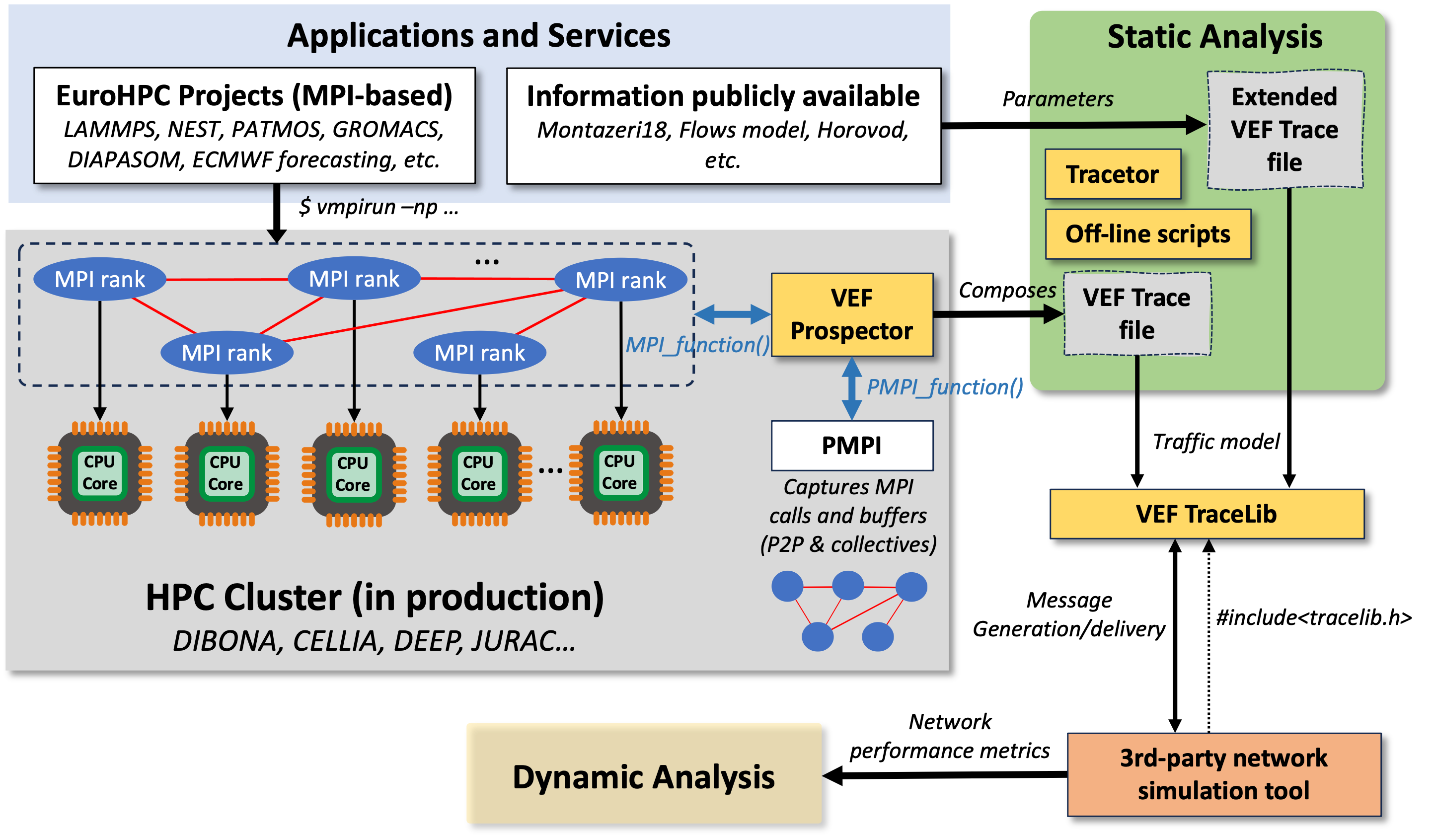}
    \caption{Diagram of our traffic-modeling methodology. Yellow text squares indicate the applications and tools of the VEF Traces framework.}
    \label{fig:vef-traces-framework}
\end{figure}

To characterize specific communication patterns, we have developed a methodology that extends and leverages the open-source framework VEF Traces~\cite{Andujar16jsc,Andujar23jsc}, consisting on i) profiling MPI-based applications communication and recording this information in network traffic traces, ii) modeling realistic network workloads considering the information of network communication patterns made publicly available by HPC systems and Data-center users, iii) reproducing these communication patterns in a specific simulation tool, and iv) performing different types of analyses (called static or dynamic) using the characterized patterns.
Figure~\ref{fig:vef-traces-framework} shows a general overview of the workflow of our proposed methodology, including the most prominent tools of the VEF Traces framework (in yellow).

In detail, the VEF Traces framework generates traffic traces that capture the communication operations of MPI-based parallel applications. To do that, the \emph{VEF Prospector} suite provides a set of libraries and tools for profiling these operations. This process requires that the target application and \emph{VEF Prospector} run concurrently on a real HPC cluster. \emph{VEF Prospector} leverages the \emph{Profiling MPI} (PMPI) library to capture the MPI calls (either point-to-point or collective) per MPI rank. The MPI call information per MPI rank is stored locally on each end node in the cluster and later combined into a single VEF trace.
Although trace profiling is a reliable way of modeling communication operations, its lack of scalability is its main weakness. Note that a real HPC cluster is required to run the application and collect the traces. Moreover, traces cannot be generated for applications with MPI task counts exceeding the number of cluster cores.

On the other hand, in data centers, there are applications other than MPI-based ones whose workloads are of interest for congestion characterization.
However, in this case, we cannot use \emph{VEF Prospector} or its profiling functionality to collect traces.
To overcome this issue, assuming profiling tools similar to PMPI are unavailable, another option is to use application-level information to characterize communication patterns, such as the flow size distribution, the source-destination distribution, and the time consumed by these patterns in the network.
Sometimes, this information is publicly available from several sources~\cite{Montazeri18sigcomm,Gonzalez19hpcs}, so it is possible to configure a set of parameters (e.g., packet destination distribution, traffic flow size distribution, number of traffic flows, generation period, traffic classes, etc.) to build a trustworthy communication model, which could eventually be scaled out to evaluate its impact on the largest interconnection networks.
The VEF traces framework also supports this approach through the ``extended'' VEF traces~\cite{Andujar23jsc}, which permits scaling out the behavior of a set of communication patterns to thousands of nodes.
This approach allows us to model the communication patterns of widely used applications in Data centers.
Examples of these patterns include those derived from online data-intensive services (e.g., those used in Facebook Data centers) or deep learning training models (e.g., those run on top of the Horovod distributed framework).

To replay the behavior recorded in VEF traces in a simulated network environment, the VEF framework provides another open-source software library, called \emph{VEF TraceLib}, which generates communication messages for a VEF trace and manages dependencies among them.
TraceLib is simulator-agnostic; it does not model network-specific behavior, which is left to the simulators. TraceLib provides a dedicated API with primitives that any simulator can use to inject messages into the network and be notified when they arrive at their destination.  There is an implementation reference of TraceLib in the VEF-Traces RED-SEA blog post\footnote{The VEF-Traces RED-SEA blog post: \href{https://redsea-project.eu/the-vef-traces-framework/}{https://redsea-project.eu/the-vef-traces-framework/}} and the \texttt{tracetor} tool\footnote{Source code of the \texttt{tracetor} tool \href{https://gitraap.i3a.info/fandujar/VEF-TraceLIB/-/blob/master/test/main.c}{https://gitraap.i3a.info/fandujar/VEF-TraceLIB/-/blob/master/test/main.c}}.
Network simulation tools that use TraceLib and VEF traces should provide a set of performance metrics to help network designers analyze communication patterns and their impact on network performance, including degradation caused by bottlenecks, reliability issues, or power consumption.
Hereafter, we refer to the analysis of these specific metrics using simulation results as \emph{dynamic analysis} (see Figure~\ref{fig:vef-traces-framework}), since the simulated network behavior and the resulting performance metrics vary over time and depend on the network configuration parameters and the nature of applications' communication operations.

This paper describes a methodology (see Figure~\ref{fig:vef-traces-framework}) to leverage the VEF Traces framework and network simulation tools to characterize the communication patterns of real applications. 
After generating the VEF traces, we analyze the communication patterns they capture in two ways. First, we have extended the VEF Prospector tool with a toolchain that helps perform a \emph{static analysis} using communication metrics, such as the operation type (P2P or collective), the MPI calls count, number of generated bytes, source and destination end-nodes engaged in the communication, etc. This toolchain generates a set of plots that can be combined into a single-file report.
For the \emph{dynamic analysis}, we have used the SAURON~\cite{Sauron} simulator, based on the well-known OMNeT++ framework~\cite{Omnetpp}. SAURON has been widely used and tested in the last decade, modeling specific, state-of-the-art interconnect technologies (e.g., InfiniBand~\cite{Yebenes16sauron-voq} or BXI~\cite{GomezLopezREGQL24,RosaGAESAL25}).
We have extended the SAURON simulator to provide metrics and plots that help analyze the impact of congestion on network performance.
The main contributions of this paper are the following:
\begin{itemize}
    \item We have studied the communication operations generated by several parallel applications widely used in Supercomputers and Data centers, such as LAMMPS, NEST, GROMACS, PATMOS, or DIAPASOM, which were run in real cluster infrastructures to record VEF traces. These traces have been uploaded to the public VEF traces repository\footnote{Repository URL: \url{https://gitraap.i3a.info/jesus.escudero/vef-traces-repository}}, so the community can use them.
    \item We have extended the VEF traces framework with additional tools that help analyze the information stored in the trace files. These tools enable us to perform a \emph{static} analysis to understand applications' traffic patterns behavior and to identify potential bottlenecks (i.e., congestion scenarios) that they may generate.
    \item We have performed the \emph{static} analysis on a set of VEF traces, analyzing the behavior of collective operations, the expected bandwidth required from the network, the source and destination distribution, etc.
    \item We have also performed a \emph{dynamic} analysis by running the aforementioned VEF traces in the SAURON network simulator, under realistic network technologies, such as InfiniBand or BXI. This analysis, based on performance metrics such as flow completion time (FCT) and network resource utilization, helps us assess the impact of network congestion on the simulated scenarios.
\end{itemize}

We expect that the proposed methodology will help network designers devise and configure efficient high-performance interconnection networks.
The paper is organized as follows.
Section~\ref{sec:case-study} describes a specific use case where we have used the proposed framework.
Section~\ref{sec:perf-ev} details the experiments and analyses (i.e., static and dynamic) that we performed on the used VEF traces.
Section~\ref{sec:related-work} provides related works on this topic.
Finally, some conclusions are drawn in Section~\ref{sec:conclusions}.

\section{Case Study: Network Workloads from EuroHPC-JU funded projects}
\label{sec:case-study}

As described before, the VEF traces framework allows MPI-based applications to be profiled using \emph{VEF Prospector}, so the MPI calls (P2P or collective) are recorded into a VEF trace file.
We can analyze the communication pattern recorded in VEF traces using the set of tools described later, and replay this trace in a network simulator.
This section describes the tasks performed using the proposed methodology (see Figure \ref{fig:vef-traces-framework}) during an international collaboration among several European R\&I projects (funded by EuroHPC-JU), such as RED-SEA\footnote{Project homepage: \href{https://redsea-project.eu/}{https://redsea-project.eu/}}, DEEP-SEA\footnote{Project homepage: \href{https://www.deep-projects.eu/}{https://www.deep-projects.eu/}}, IO-SEA\footnote{Project homepage: \href{https://iosea-project.eu/}{https://iosea-project.eu/}}, and MAELSTROM\footnote{Project homepage: \href{https://www.maelstrom-eurohpc.eu/}{https://www.maelstrom-eurohpc.eu/}}.
This cross-collaboration has permitted several partners to gather VEF traces from HPC applications and benchmarks, such as NEST~\cite{NEST}, GROMACS~\cite{GROMACS}, LAMMPS~\cite{LAMMPS}, PATMOS~\cite{PATMOS}, DIAPASOM~\cite{DIAPASOM}, and Weather Forecasts (based on machine learning)~\cite{ecmwf}.
These traces have been uploaded to a public repository~\cite{VEF-traces-repository}, so they can be used to feed network simulators whenever they use the \texttt{TraceLib} library API.
In the following, we focus on four representative applications (NEST, GROMACS, LAMMPS, and PATMOS) and describe their computational models, domain decomposition, and parallelization strategy. We also detail how the static and dynamic analyses enabled by the VEF framework have been performed on the VEF traces generated from different runs of these applications.

\subsection{Applications description}

\subsubsection{NEST}
\label{sec:case-study:NEST}

The \emph{NEural Simulation Tool} (NEST)~\cite{NEST} models spiking neural networks of any size with focus on the dynamics, size, and structure of neural systems rather than on the exact morphology of individual neurons. Examples of these models include studies on spike-timing, dependent plasticity in extensive simulations of cortical networks, verification of mean-field models, and models of Alzheimer's disease, Parkinson's disease, and tinnitus. To simulate these models, NEST defines a neural system (also called the NEST network) composed of a potentially large number of neurons and their connections.
In a NEST network, different neuron and synapse models can coexist. Any two neurons can have multiple connections with several properties. Thus, in general, connectivity cannot be described by a weight or connectivity matrix, but rather by an adjacency list.
The NEST computational model uses a hybrid parallelism approach that combines shared- and distributed-memory architectures.
NEST assigns neurons to \emph{virtual processes} (i.e., MPI tasks or OpenMP processes) that compute the evolution of a corresponding fraction of the total number of neurons.

Specifically, NEST uses OpenMP for processes on a single node that communicate ``all-to-all'' using shared memory.
The MPI processes are mapped to different nodes in the computing cluster, and they communicate all-to-all in a distributed-memory architecture. Mapping OpenMP and MPI processes to computing cluster resources is left to NEST users and the cluster's scheduling policies.
This extensive communication between OpenMP and MPI processes requires distributing the different NEST model fractions and synchronizing them to maintain consistency.

Therefore, each MPI process keeps its status synchronized with the other processes, exchanging data via extensive \texttt{MPI\_Alltoall} calls after each model-integration step.
Moreover, calls to \texttt{MPI\_AllReduce} and \texttt{MPI\_Barrier} are required for synchronization.
These collective operations may be prone to generating congestion scenarios, as we show later.

\subsubsection{GROMACS}
\label{sec:case-study:GROMACS}

The \emph{GROningen MAchine for Chemical Simulations} (GROMACS)~\cite{GROMACS} is one of the most widely used open-source software codes in chemistry.
It performs molecular dynamics simulations using Newtonian equations of motion for systems with hundreds to millions of particles.
GROMACS is primarily designed for biochemical molecules such as proteins, lipids, and nucleic acids, which involve numerous complex bonded and non-bonded interactions.
In addition to GROMACS, other options for molecular dynamics (MD) simulations include LAMMPS, as described later.

Typically, an MD user selects an initial molecular configuration, specifies the atomic interactions and model physics, runs a simulation, and observes the trajectory. Such simulations evaluate the millions of particle interactions over billions of time steps, which can require extraordinary amounts of computational hardware and time.
GROMACS exploits MPI to distribute the simulation across computing nodes, OpenMP for multi-threading within each node, and vector extensions on CPUs (AVX2) and GPUs to achieve maximum efficiency.

GROMACS splits the problem into independent work units and distributes them across ensembles of simulations, multiple program paths, and domains. Spatial domain decomposition efficiently partitions the simulation to preserve reference locality within each domain.
Specifically, a spatial domain is assigned to each rank, and the equations of motion for particles within its local domain are integrated.
This data parallelization maps each domain to an MPI rank, which can access CPUs and GPUs within each cluster computing node.

In general, GROMACS has been thoroughly optimized to reduce communication among MPI ranks. However, model scalability threatens this principle because interatomic interactions (short- and long-range) scale with the number of atoms. So it increases the volume of all-to-all communication among the MPI ranks. Although this communication involves multiple MPI collective operations, \texttt{MPI\_Broadcast} and \texttt{MPI\_AlltoAll} dominate the amount of exchanged information, which may stress the computing cluster interconnection network.

\subsubsection{LAMMPS}
\label{sec:case-study:LAMMPS}

The \emph{Large-scale Atomic/Molecular Massively Parallel Simulator} (LAMMPS)~\cite{LAMMPS} is a classical molecular dynamics (MD) code with a focus on materials modeling\footnote{https://www.lammps.org/}, such as solid-state materials (metals, semiconductors), soft matter (biomolecules, polymers), and coarse-grained or mesoscopic systems. It can be used to model atoms or, more generally, to serve as a parallel particle simulator at the atomic, meso, or continuum scale. It is also well known for its ease of compiling and running on several different computer architectures (from laptops to large clusters).
As the model's computational cost scales linearly with the number of simulated atoms, LAMMPS is designed for distributed-memory MPI-based parallelism, so with a few hundred atoms/core, most of its models are scalable to millions of CPU cores.

Specifically, LAMMPS defines a 2-D or 3-D simulation box to model the interaction of atoms (or particles), then \emph{partitioning} this box into variable-sized and non-overlapping subdomains containing the same number of atoms. These subdomains are distributed across the cluster's memory and assigned to different processors. Each processor stores information (positions, velocities, etc.) for
the subset of (\emph{own}) atoms within its subdomain, and stores copies of some of that information for other subdomain atoms (i.e., \emph{ghost}) within a cutoff distance. This enables each processor to calculate short-range interactions that involve its own and ghost atoms.
However, copies of ghost atoms must be updated, which requires communication with the processors that own them during long-range interactions among atoms.

This information exchange among processors defines the communication pattern of LAMMPS that contains point-to-point (P2P) communications to transfer information to nearby domains, plus collective operations (mostly \texttt{MPI\_Broadcast} and \texttt{MPI\_AllReduce}) to update the domain population, and \texttt{MPI\_AlltoAll} to perform fast-fourier transform (FFT) operations involving long-range atom interactions. As we demonstrate later, based on static analysis of a LAMMPS VEF trace, communication is dominated by the \texttt{MPI\_AllReduce} collective operation, which can cause network congestion.

\subsubsection{PATMOS}
\label{sec:case-study:PATMOS}

It is a Monte Carlo neutron transport code developed at CEA (France) that prototypes applications for nuclear safety and radiation shielding~\cite{PATMOS}.
Its final goal is to perform pin-by-pin full-core depletion calculations for large nuclear power reactors with realistic temperature fields.
PATMOS has been conceived to support two levels of parallelism (i.e., hybrid parallelism). The first level corresponds to distributed-memory parallelism and relies on MPI. The second level corresponds to shared-memory parallelism using OpenMP or POSIX threads.

In PATMOS, simulations are divided into batches, even if all source particles are independent (for instance, one million histories are typically simulated as 1000 batches of 1000 particles each). Because of their size, the only non-duplicate mutable structures are the scores, which are concurrently modified as shared objects via OpenMP atomic adds.
PATMOS is multi-threaded, so particles in the batch/cycle are dispatched in equal numbers to available threads, which simulate the particle history and compute their contributions to the score. This parallelization procedure is deterministic, thus assuring reproducibility and facilitating debugging. It is possible to use dynamic dispatch of particles to available threads to ensure load balancing, but in this case, the result is non-deterministic. Each thread has its own random number generator, initialized independently at the start of the simulation.

Each MPI process executes an independent simulation. At the end of the run, the results of all simulations are aggregated via MPI reduction operations (i.e., \texttt{MPI\_AllReduce} for computing the mean, followed by a call to \texttt{MPI\_Reduce} for the variance calculation).

\subsection{Static analysis}
\label{sec:case-study:static-analysis}

As mentioned above, VEF traces can be analyzed offline before their use in a network simulator.
\emph{VEF prospector} provides two tools for this purpose:

\begin{itemize}
    \item \texttt{\textbf{tracetor}} is a simple application written in the C language to validate a VEF trace after it was generated with VEF-Prospector. It is included in the VEF-TraceLib repository under the \texttt{test/} directory as a simulator-independent implementation reference. This application uses the \texttt{TraceLib} library to execute all the communication operations recorded in the trace and insert them in an ideal network (i.e., a bus with infinite bandwidth). In this way, \texttt{tracetor} analyzes the number of communication operations (and the messages in which they are decomposed) at first glance, before reproducing them in a simulation environment. Most importantly, \texttt{tracetor} also enables testing that all messages are generated and received correctly and that there are no failures in the dependencies between messages recorded in the VEF trace file.

    \item \textbf{offline-vef-analysis.sh} is a script that characterizes the VEF trace by providing an extensive number of plots, text files, and PDF reports, with specific information about message generation, destination generation distribution, workload size, number, and types of collective operations, etc., which are gathered in a VEF trace generated through VEF-Prospector. 
\end{itemize}

For MPI collective operations, we assume the $N$ delivery algorithm in \texttt{TraceLib}. That is, for one-to-many collectives (e.g., \texttt{MPI\_Broadcast} or \texttt{MPI\_Scatter}), the root task generates one message for each remaining task in the communicator. For instance, for a \texttt{MPI\_Broadcast} operation on a communicator with $N$ tasks, the total bytes sent by the root task in bytes is $N-1~\times~{size}_{bcast}$ bytes. Subsequently, many-to-one collectives (e.g., \texttt{MPI\_Reduce} or \texttt{MPI\_Gather}) generate as many messages as the communicator has non-root tasks. For point-to-point operations, applications select tasks at runtime. In any case, our analysis tools do not alter the applications' communication patterns.

Using these tools, we performed static analyses on all the traces obtained. We uploaded one report per VEF trace to the already-mentioned public repository.
Although static analysis provides valuable information, it is insufficient to explain how this traffic will behave under specific network configurations or the congestion scenarios it may generate.
For this reason, we need additional insights from network simulation tools that use VEF traces and analyze specific metrics to characterize congestion, i.e., \emph{dynamic analysis}.

\subsection{Dynamic analysis}

To replay VEF trace behavior in a network simulation tool and perform the \emph{dynamic analysis}, we first need to integrate the simulator with the \texttt{TraceLib} library. The \texttt{TraceLib} library API provides functions for reading VEF traces, mapping tasks to end nodes, managing trace execution, and communicating with the simulator, so the network simulator only needs to ask the library to supply the application messages recorded in a VEF trace. For each obtained message, the network simulator splits it into packets according to the network MTU, injects these packets into the network, models how these packets are transferred from a source end node to a destination end node, composes messages again when packets arrive at the receiving node, and returns these messages to \texttt{TraceLib}, which finally processes them.
Depending on the dependencies among messages, as recorded in a VEF trace, message reception at \texttt{TraceLib} may automatically generate dependent messages.

For each generated message, \texttt{TraceLib} provides the simulator with the source and destination end nodes, the message length, and the message ID (used to uniquely identify the message within \texttt{TraceLib}). 
Note that the simulator does not need to distinguish whether a specific message was generated due to a point-to-point or collective communication operation, or whether an application task is stopped waiting for messages that the simulator has not yet provided to \texttt{TraceLib}.
Moreover, \texttt{TraceLib} includes features for simulator developers, such as the ability to simulate multiple, even simultaneous, traces, a flexible mapping scheme for MPI tasks to end-nodes, and the ability to implement specific collective communication functions (e.g., those implemented by OpenMPI or MVAPICH). It is also possible to have idle end-nodes and end-nodes with multiple tasks allocated (even tasks from different traces). It is worth noting that MPI collective communication has been implemented in \texttt{TraceLib} based on the algorithms available in the OpenMPI driver, and these algorithms can be extended and improved by the community.

In our proposed framework, \texttt{TraceLib} has been integrated with the OMNeT++-based SAURON simulator~\cite{Sauron}. Note that the network simulator has also been extended to provide multiple metrics for dynamic analysis of VEF trace behavior across the configured network scenarios.
SAURON models communication operations at a packet level, meaning it performs message packetization.
This feature allows us to observe the evolution of traffic patterns over time in the VEF trace, using key network metrics such as application execution time, FCT, and switch queue occupancy.
We have also added a new feature to the SAURON simulator: a GUI that colors network links based on buffer occupancy.
This coloring can be observed ``over time'' for a specific VEF trace run in the simulator.
This feature, combined with the static and dynamic analyses, can be used to characterize the network congestion.

\section{Performance Evaluation}
\label{sec:perf-ev}

Based on the methodology described previously (see Figure~\ref{fig:vef-traces-framework}), this section presents simulation results from experiments using the previously mentioned applications. These experiments allow us to characterize realistic communication patterns and congestion scenarios.
In the following sections, we describe the experiment configuration, static analysis, and dynamic analysis of these traces using the SAURON network simulator.

\subsection{Experiments configuration}
\label{sec:exp_conf}

We have selected VEF traces from NEST, GROMACS, LAMMPS, and PATMOS applications, available in the VEF traces repository~\cite{VEF-traces-repository} (see Section~\ref{sec:case-study}).
Specifically, we have chosen VEF traces configured with $64$ and $256$ MPI ranks.
We have performed the \emph{static analysis} of these traces, examining several metrics such as the number of collective calls per operation, the amount of traffic generated by these operations, the message and byte count versus execution time, and the source/destination message and number of bytes exchanged.
These results are shown in Section~\ref{sec:ev:static}.
Next, we used these VEF traces to feed the SAURON simulator.
These results can be seen in Section~\ref{sec:ev:dynamic}.
In more detail, we have configured SAURON to model realistic network technologies.
Table~\ref{tab:netconfs} shows two configurations for the network architecture.

\begin{table}[!htb]
\centering
\resizebox{.8\columnwidth}{!}{%
\begin{tabular}{lcc}
\hline
\textbf{Network parameters} & \textbf{Configuration \#1} & \textbf{Configuration \#2} \\ \hline
Switch architecture  & Combined Input-Output Queues (CIOQ)~\cite{CIOQ} & Input-Queued (IQ)~\cite{karol1987input} \\
Input buffer size (KiB)  & 128 & 128\\
Output buffer size (KiB) & 48 & - \\
Flow control & Priority Flow Control (PFC)~\cite{pfc8021qbb}  & Credit-based~\cite{CreditFC} \\
MTU (Bytes)  & 9600 (JUMBO frames) & 4096\\
Variable packet size & Yes & No\\
Link bandwidth (Gbps)  & 400 & 100\\
Link length (m) & 3    & 3 \\
Link latency (ns/m) & 5 & 5 \\
Number virtual channels  & 1 & 1\\
\hline
\end{tabular}%
}
\caption{Network configurations used in the Sauron simulator.}
\label{tab:netconfs}
\end{table}

Network configuration~$\#1$ has features similar to high-speed Ethernet networks, InfiniBand, or next-generation BXI, while network configuration~$\#2$ defines features used in network devices from previous generations (e.g., $100$ Gbps link bandwidth), which are still utilized in some HPC systems and Data centers today.
It is worth mentioning that our objective is to understand how communication operations recorded in VEF traces affect different network configurations, rather than to demonstrate which configuration is faster.

\begin{figure}[!htb]
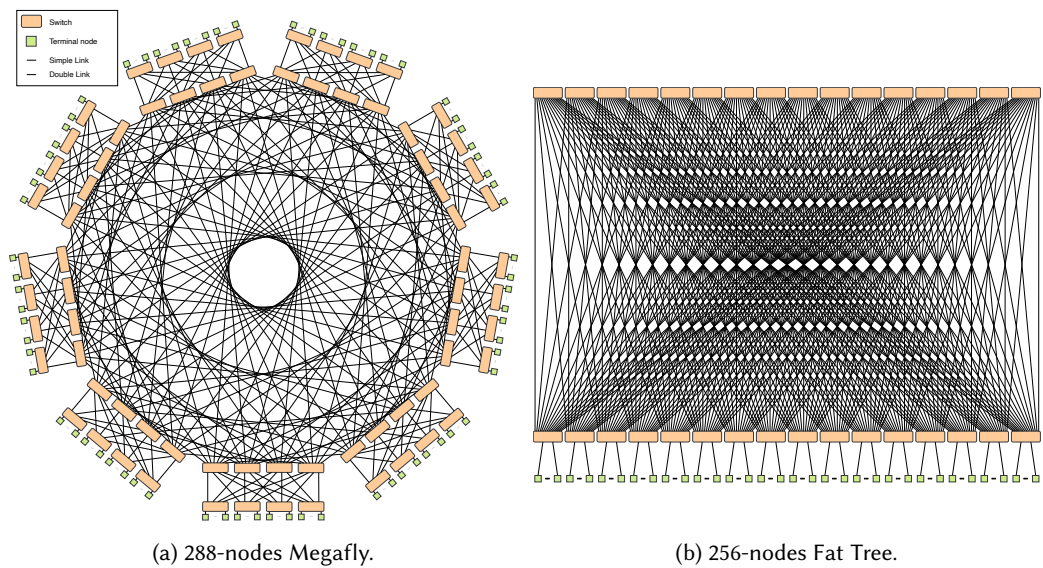

    \centering
	\subfloat[288-nodes Megafly.]{
		\includegraphics[width=0.45\textwidth]{figures/288-nodes-MegaFly.pdf}
		\label{fig:288-nodes-Megafly}
	}
	\subfloat[256-nodes Fat Tree.]{
        \raisebox{0.6cm}{\includegraphics[width=0.45\textwidth]{figures/256-nodes-FatTree.pdf}}		
        
        \label{fig:256-nodes-FatTree}
	}
\caption{Topologies used for the dynamic analysis experiments.}
 \label{fig:Topologies}
\end{figure}

Table~\ref{tab:TopologiesCharacts} shows the network topologies used in our dynamic analysis: a $256$-node Fat-tree and a $288$-node Megafly~\cite{flajslik2018megafly}.
These topologies are illustrated in Figure~\ref{fig:Topologies}. 
We only show two terminal nodes connected to each leaf switch for clarity. 

\begin{table}[!htb]
\centering
\resizebox{.4\columnwidth}{!}{%
\small
\begin{tabular}{lcc} 
\cline{2-3}
      & \textbf{Fat Tree} & \textbf{Megafly}  \\ 
\hline
Number of~ terminal nodes & 256      & 288         \\ 
\hline
\begin{tabular}[c]{@{}l@{}}Number of terminal nodes \\in each leaf switch\end{tabular}  & 16       & 8           \\ 
\hline
Number of leaf switches & 16       & 36          \\ 
\hline
Switch radix       & 32       & 16          \\ 
\hline
Number of Switches  & 32       & 72          \\ 
\hline
Network diameter    & 3        & 4           \\ 
\hline
Number of switch ports used & 768  & 1152        \\
\hline
\end{tabular}%
}
\caption{Network topologies configuration parameters.}
\label{tab:TopologiesCharacts}
\end{table}

%

Note that the Fat Tree is widely used in HPC systems and data centers. However, new topologies based on Dragonfly~\cite{kim_technology-driven_2008} ones have been proposed, providing better scalability as the number of nodes increases. One problem with Dragonfly networks is that the minimum-length path diversity is lower than that of Fat Trees.
Therefore, Megafly~\cite{flajslik2018megafly} (a.k.a.~Dragonfly+~\cite{shpiner_dragonfly_2017}) topologies have been proposed to overcome this issue.
As we can see in Figure~\ref{fig:288-nodes-Megafly}, the number of paths between two terminal nodes in different groups is multiple, compared to that in a Fat tree (see Figure~\ref{fig:256-nodes-FatTree}).

Regarding the routing algorithm, we have configured the network to use $D$-mod-$K$ deterministic routing~\cite{zahavi12dmodk}, which efficiently balances traffic across available paths in the topology, making it a good option for all-to-all communication.
Note that this deterministic routing algorithm performs similarly to adaptive routing algorithms under certain traffic patterns~\cite{gomez2015destro}.
Unfortunately, other traffic patterns generated by collective communication in the network, such as all-to-one or all-to-many, may cause congestion, degrading routing performance.
Note that the study of other routing algorithms aimed at alleviating congestion problems, such as oblivious or adaptive, is out of the scope of this paper.

The SAURON simulator provides a range of metrics to evaluate the performance of the configured networks. Among the previously defined scenarios, the most useful for dynamic analysis are the execution (i.e., simulated) time of a VEF trace and the Flow Completion Time (FCT), which measures the average time from when an application message is generated until it is completely received at the destination end node.
SAURON also provides metrics to analyze switch buffer occupancy over time, helping identify communication bottlenecks and characterize network congestion.

\subsection{Static analysis results}
\label{sec:ev:static}

The main objective of the \emph{static analysis} is to study the behavior of a VEF Trace before it is used in a simulated environment.
This analysis is intended to show the number of messages exchanged among end nodes, the source/destination distribution, expected bytes exchanged, etc., so it does not assume a specific network topology or implementation for collective operations message exchange.
This section presents results obtained using the scripts described in Section~\ref{sec:case-study:static-analysis} on VEF traces from NEST, GROMACS, LAMMPS, and PATMOS applications.

\begin{figure}[!htb]
	\centering
	\subfloat[Number of collective calls for each operation.]{
		\includegraphics[width=.495\columnwidth]{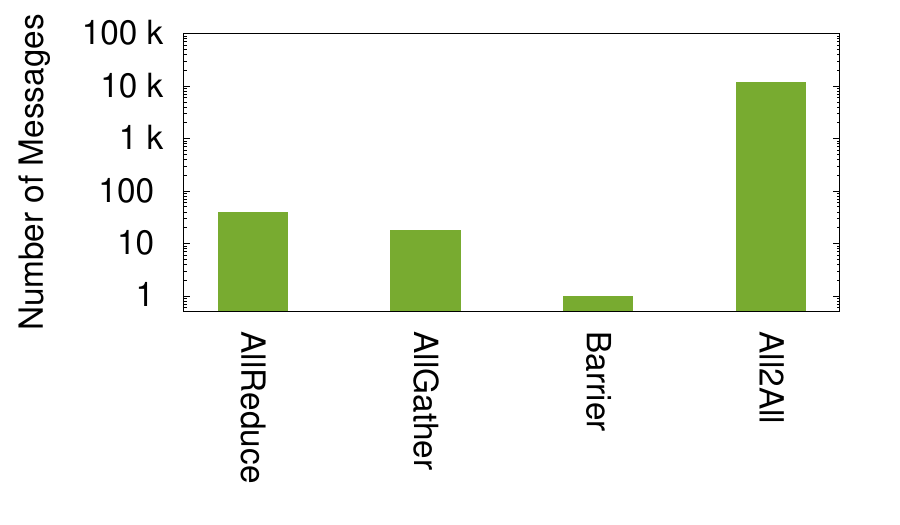}
		\label{fig:NEST.num}
	}
	\subfloat[Traffic generated by each operation.]{
		\includegraphics[width=.495\columnwidth]{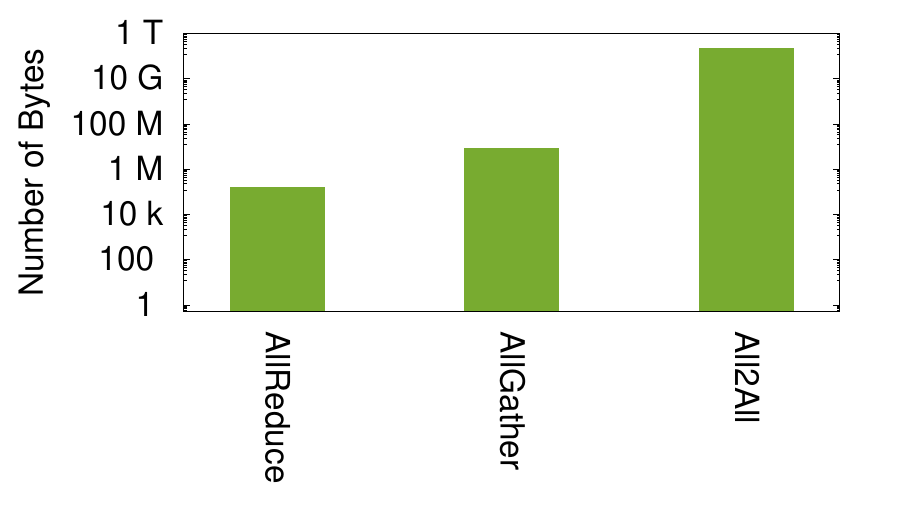}
		\label{fig:NEST.bnum}
	}

    \includegraphics[width=.75\columnwidth]{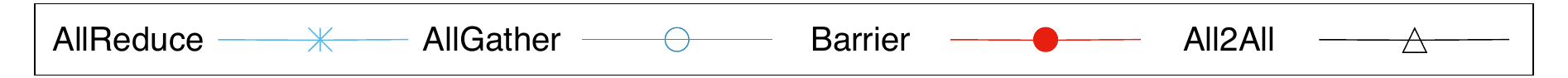}
	\subfloat[Collective operation message count.]{
		\includegraphics[width=.495\columnwidth]{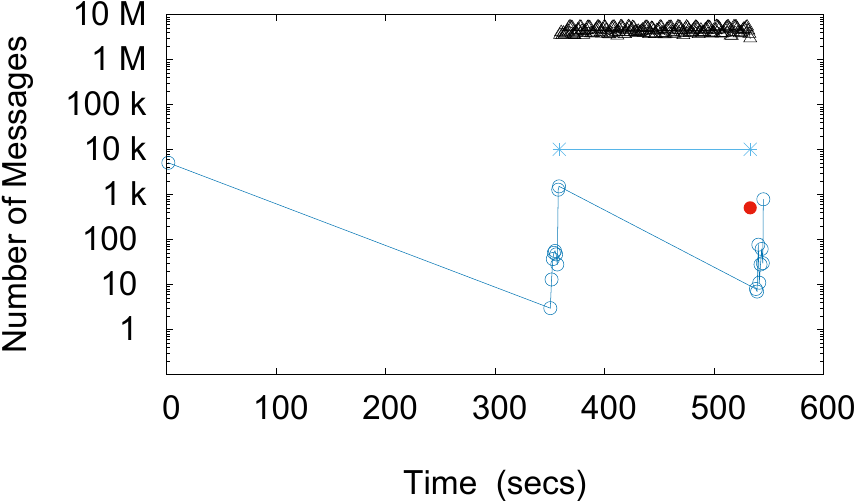}
		\label{fig:NEST-Coll-msg.num}
	}
	\subfloat[Collective operation byte count.]{
		\includegraphics[width=.495\columnwidth]{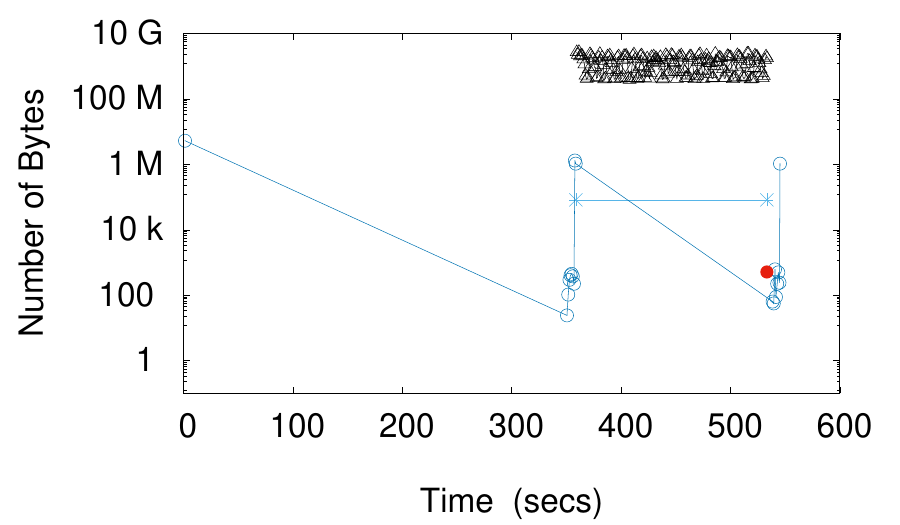}
		\label{fig:NEST-Coll-bytes.num}
	}
 
 	\subfloat[Number of exchanged messages.]{
 		\includegraphics[width=.4\columnwidth]{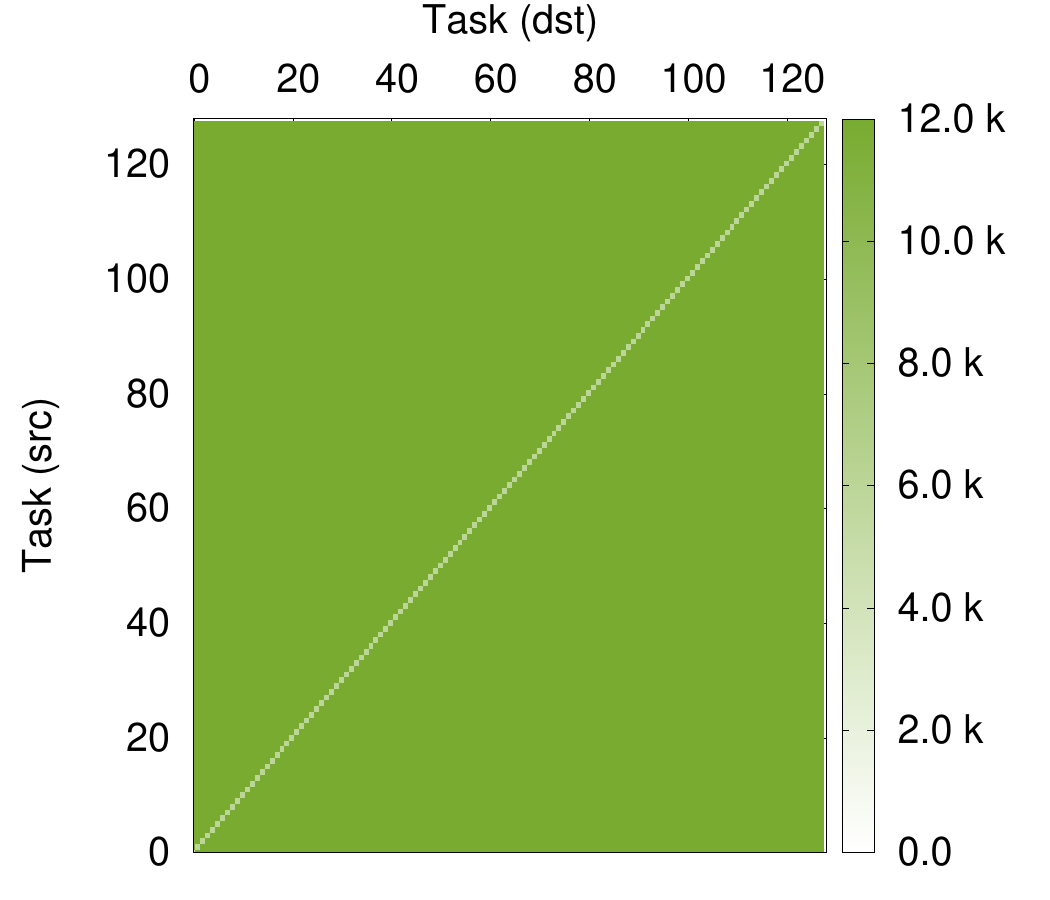}
 		\label{fig:NEST-mat}
 	}
 	\subfloat[Number of exchanged bytes.]{
 		\includegraphics[width=.4\columnwidth]{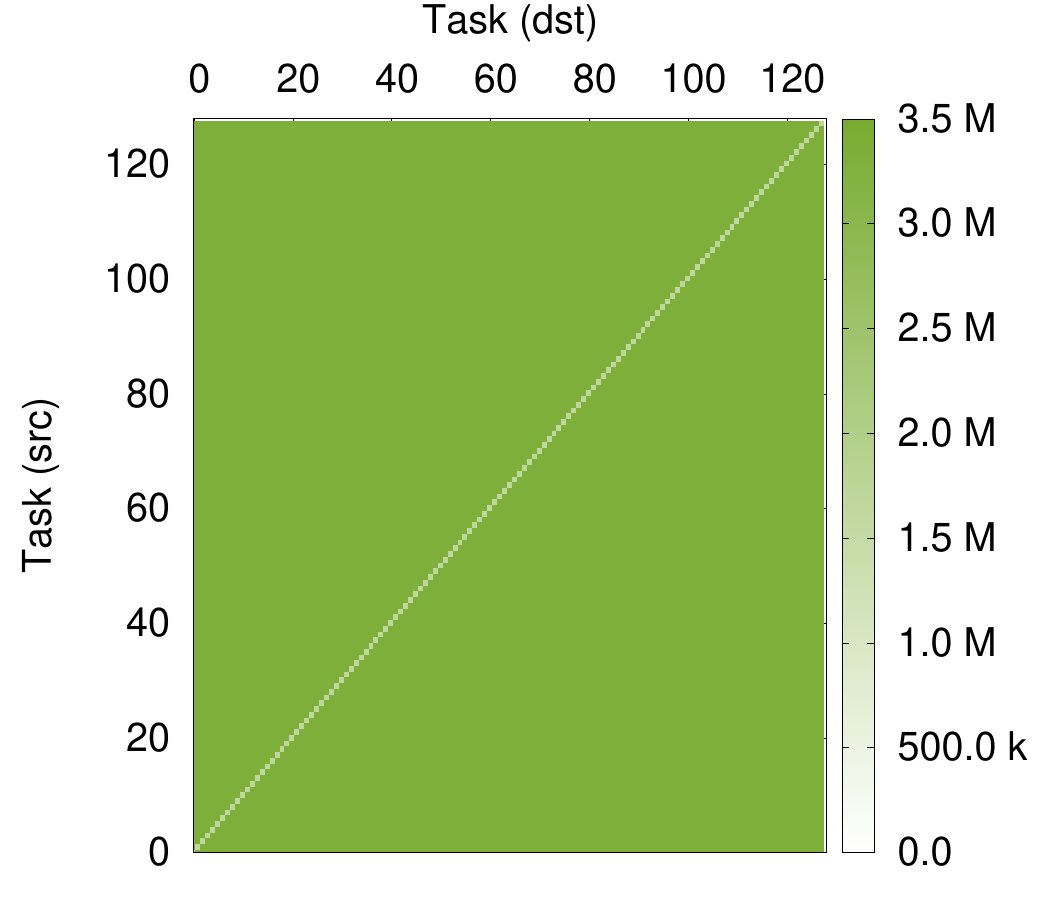}
 		\label{fig:NESTB-bmat}
 	}
 	
 	\caption{NEST static analysis results.} 
 	\label{fig:NESTstatic}
\end{figure}

Figure~\ref{fig:NESTstatic} shows the static analysis results for the NEST application.
Figures~\ref{fig:NEST.num}~and~\ref{fig:NEST.bnum} depict the number of messages and bytes generated in the network per communication operation, which provides information about which operations certainly dominate the network communication. The byte counts in the figures reflect the MPI buffer sizes when calling the collective or P2P operations. This means that, for example, a \texttt{Broadcast} operation has $s_B$ bytes sent by the root task, while the others ($size_{comm}$) receive $s_B$ bytes. Of course, the real traffic generated would be $s_B \times (size_{comm} -1)$ bytes if employing the $N$ delivery algorithm, or $s_B \times log_{(size_{comm} -1)}$ bytes if using the logN algorithm.
As we can see in Figure~\ref{fig:NEST.num}, out of the four collective operations found in the trace, \texttt{All2All} is the most predominant in the number of calls, which is three orders of magnitude larger than that of \texttt{AllReduce}. Figure~\ref{fig:NEST.bnum} shows that the \texttt{All2All} contribution to traffic in the number of bytes is more than five orders of magnitude higher than the second one. Also, the number of bytes generated by \texttt{AllGather} operations exceeds that of \texttt{AllReduce}.

Figures~\ref{fig:NEST-Coll-msg.num}~and~\ref{fig:NEST-Coll-bytes.num} show the number of messages and bytes generated in the network as a function of time, which permits identifying when a bottleneck may appear during the application execution, and which specific communication operation type is responsible for creating this situation.
As we can see in Figure~\ref{fig:NEST-Coll-msg.num}, the application starts with a burst of \texttt{AllGather} operations, until it reaches around $350$ seconds of runtime, where \texttt{AllGather} and \texttt{AllReduce} calls precede a period dominated by \texttt{All2All} operations; after that period, the application ends.
In Figure~\ref{fig:NEST-Coll-bytes.num}, the traffic generation in the number of bytes follows the trend observed in the collective operation calls. However, as noted earlier, \texttt{AllGather} operations generate more traffic with fewer calls than \texttt{AllReduce}, as shown before and after the period when \texttt{All2All} is predominant.
Note that the \texttt{All2All} traffic burst after $350$ seconds of runtime is combined with traffic of \texttt{AllReduce} and \texttt{AllGather}, so traffic flows from the latter may suffer extra latency due to congestion.

Figures~\ref{fig:NEST-mat}~and~\ref{fig:NESTB-bmat} depict the number of messages and bytes generated in the network as matrices showing source-to-destination MPI rank pairs, which enables identification of which source-destination communications are most popular and may create congested areas.
In Figure~\ref{fig:NEST-mat}, we can see that overall message exchange among MPI ranks is aligned with the operations found in the trace, where all-to-all communications are the predominant ones. This figure also shows that NEST communication flows uniformly to and from every rank. 
Note that this behavior corresponds to the NEST computation model, described in Section~\ref{sec:case-study:NEST}, where the model neurons are distributed among the application processes and communicate all-to-all (i.e., using \texttt{All2All} collective calls) their status to synchronize the data distributed throughout the system.

\begin{figure}[!hb]
	\centering
	\subfloat[Number of collective calls for each operation.]{
		\includegraphics[width=.495\columnwidth]{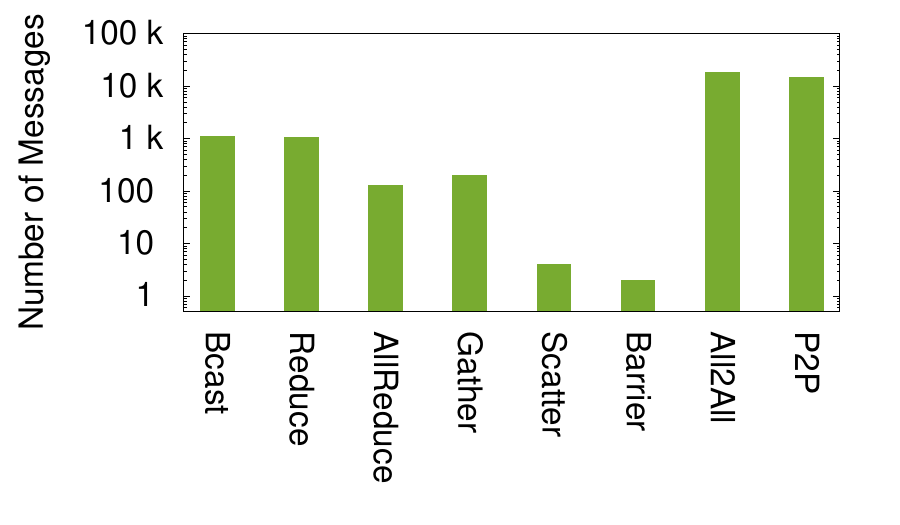}
		\label{fig:GROMACS.num}
	}
	\subfloat[Traffic generated by each operation.]{
		\includegraphics[width=.495\columnwidth]{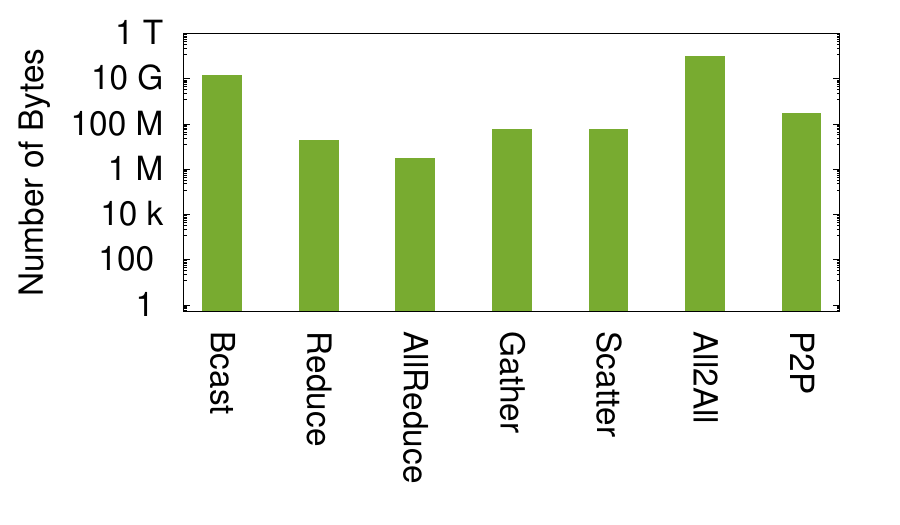}
		\label{fig:GROMACS.bnum}
	}
    
    \includegraphics[width=.75\columnwidth]{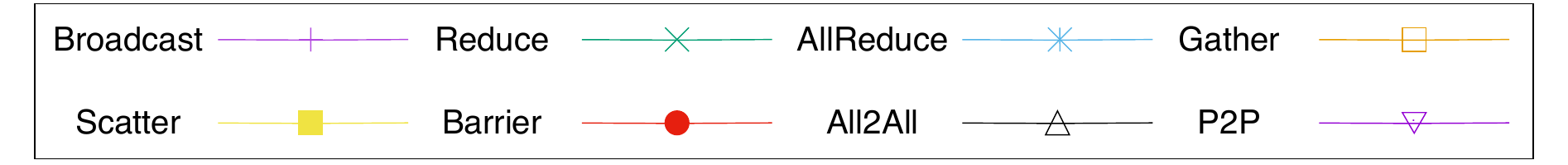}
    
    \subfloat[Collective operation message count.]{
		\includegraphics[width=.495\columnwidth]{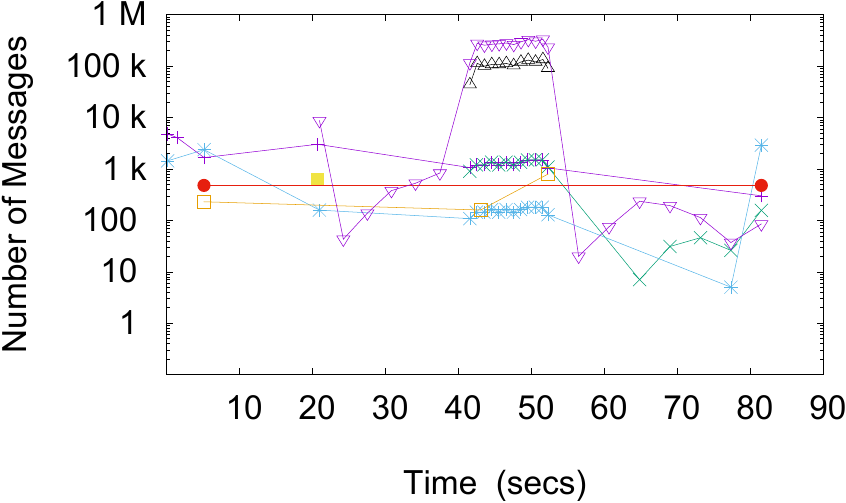}
		\label{fig:GROMACS-Coll-msg.num}
	}
	\subfloat[Collective operation byte count.]{
		\includegraphics[width=.495\columnwidth]{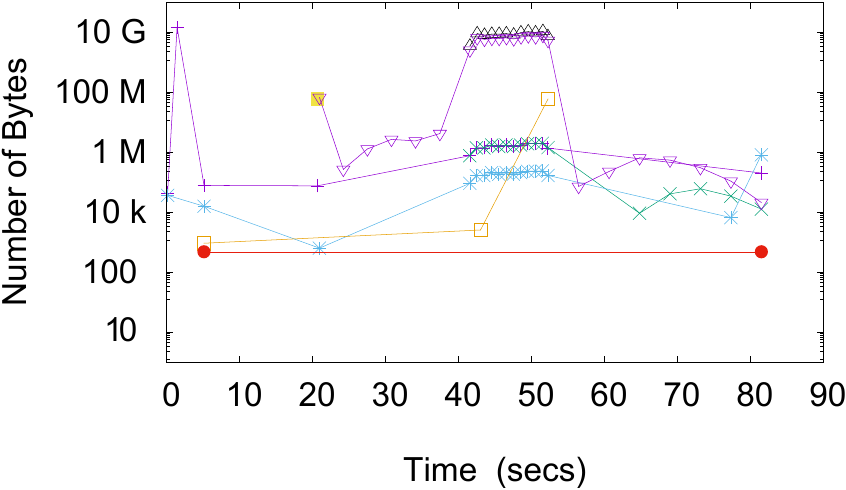}
		\label{fig:GROMACS-Coll-bytes.num}
	}

 	\subfloat[Number of exchanged messages.]{
 		\includegraphics[width=.4\columnwidth]{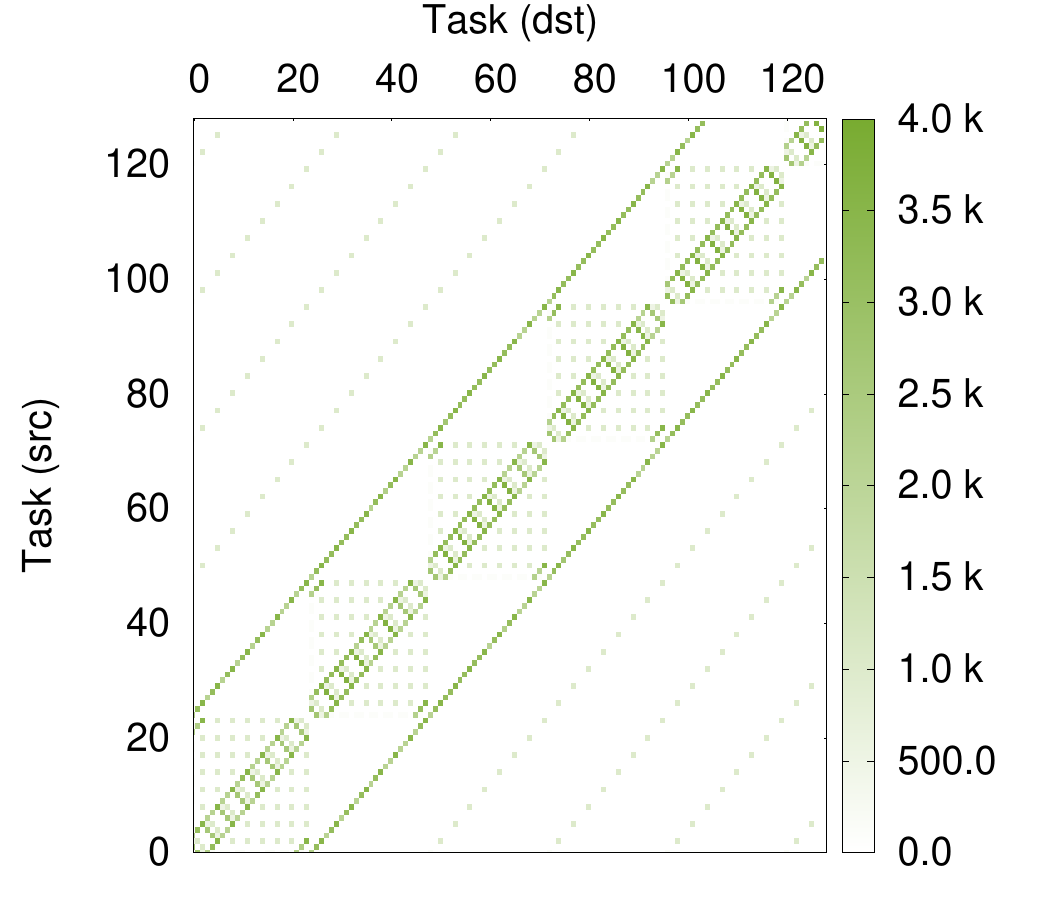}
 		\label{fig:GROMACS-mat}
 	}
 	\subfloat[Number of exchanged bytes.]{
 		\includegraphics[width=.4\columnwidth]{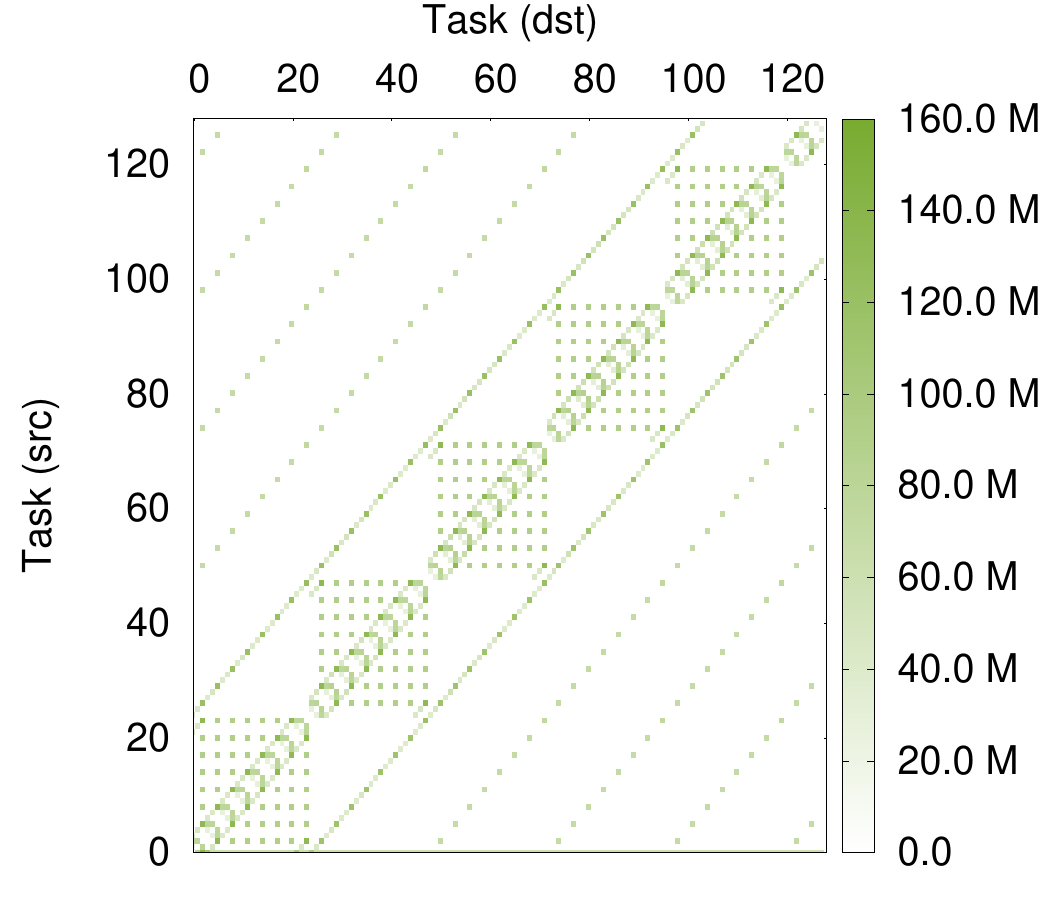}
 		\label{fig:GROMACSB-bmat}
 	}
 	
 	\caption{GROMACS static analysis results.} 
 	\label{fig:GROMACSstatic}
\end{figure}

Figure~\ref{fig:GROMACSstatic} shows the static analysis results for the GROMACS application, organized in subfigures, as we have described for the NEST application. 

Specifically, Figures~\ref{fig:GROMACS.num}~and~\ref{fig:GROMACS.bnum} show that \texttt{All2All} collective dominates the number of operation calls and generated bytes, followed by the \texttt{Broadcast} (one order of magnitude less) and \texttt{P2P} (almost three orders of magnitude less).
Figures~\ref{fig:GROMACS-Coll-msg.num}~and~\ref{fig:GROMACS-Coll-bytes.num} show the number of generated messages and bytes per communication operation.
We can see that the application calls \texttt{Broadcast} and \texttt{AllReduce} at the beginning, and there are short bursts of \texttt{AllReduce}, \texttt{Broadcast}, and \texttt{Scatter} until past 20 seconds, where \texttt{P2P} traffic is exchanged.
After 40 seconds, \texttt{All2All} and \texttt{P2P} calls generate most of the network traffic and last for around 10 seconds, so congestion may occur during this period.
After that, some \texttt{Reduce} and \texttt{AllReduce} bursts mark the end of the simulation.

Figures~\ref{fig:GROMACS-mat}~and~\ref{fig:GROMACSB-bmat} show the message and bytes exchanged among the MPI tasks, where we can see that communication occurs within groups of nodes in the network. 
This is because, as described in Section~\ref{sec:case-study:GROMACS}, GROMACS splits the problem into independent units of work and distributes them across ensembles of simulations. 
Moreover, the spatial domain decomposition partitions the application into multiple MPI communicators to group tasks, so all-to-all and broadcast communication can be restricted to exploit locality and reduce communication.

\begin{figure}[!htb]
	\centering
	\subfloat[Number of collective calls for each operation.]{
		\includegraphics[width=.495\columnwidth]{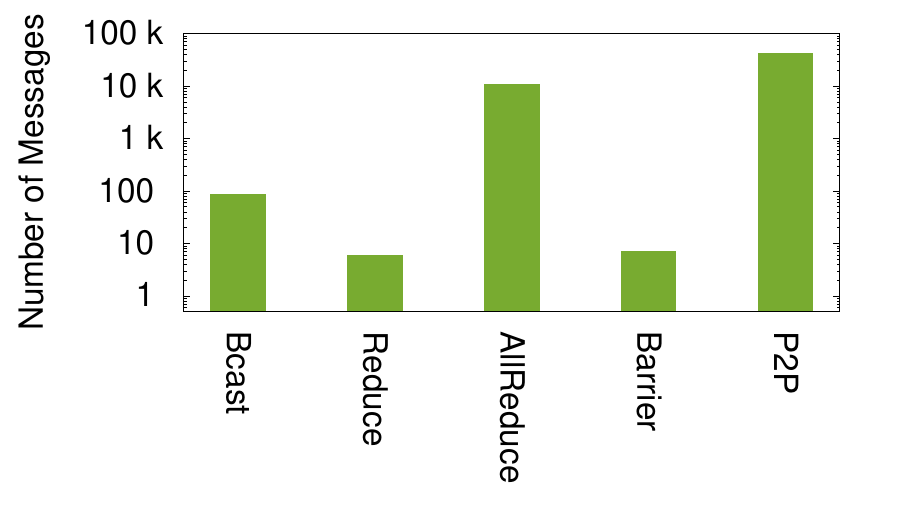}
		\label{fig:LAMMPS.num}
	}
	\subfloat[Traffic generated by each operation.]{
		\includegraphics[width=.495\columnwidth]{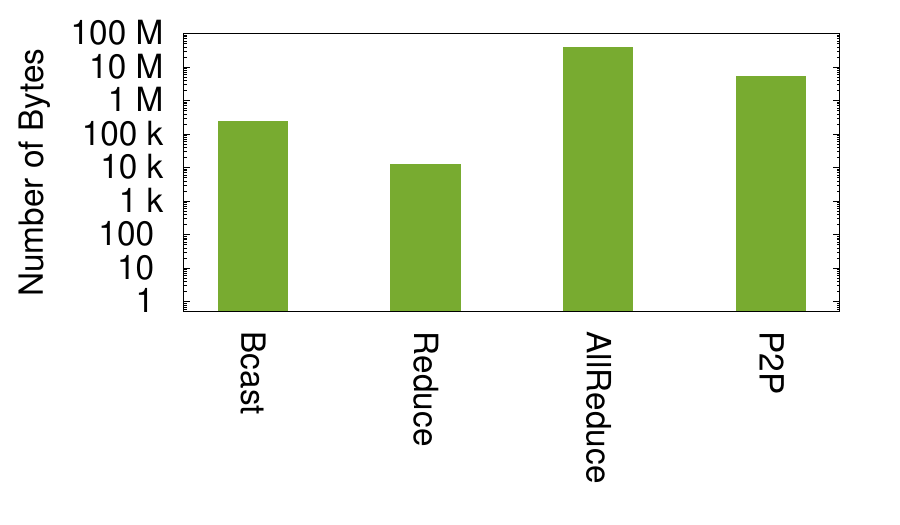}
		\label{fig:LAMMPS.bnum}
	}

    \includegraphics[width=.65\columnwidth]{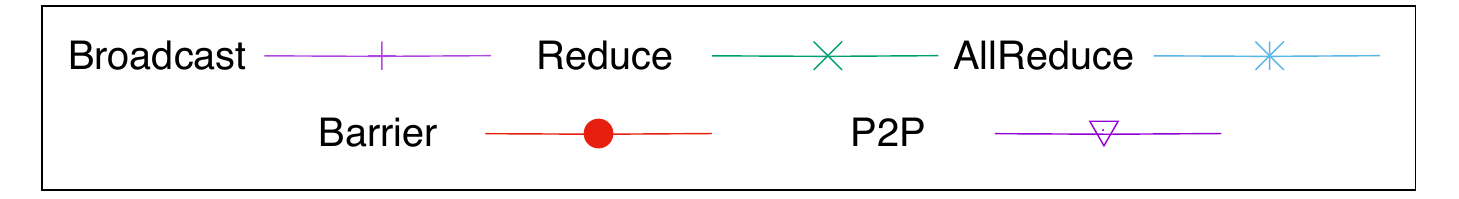}
	\subfloat[Collective operation message count.]{
		\includegraphics[width=.495\columnwidth]{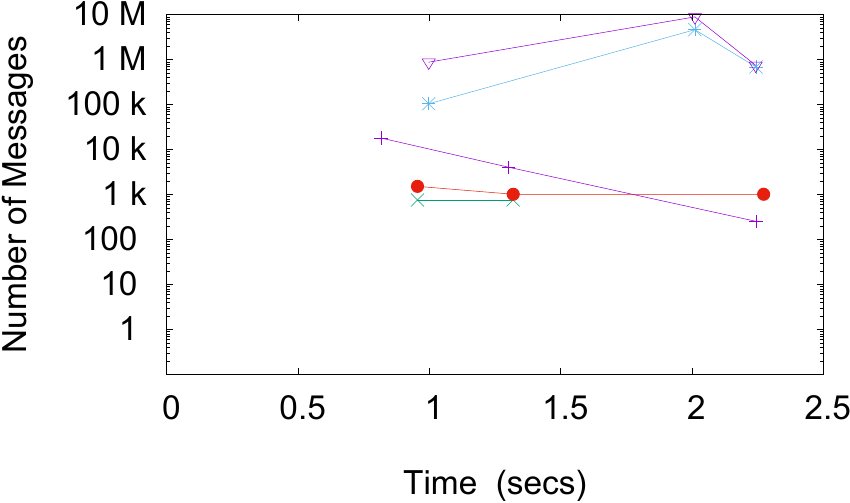}
		\label{fig:LAMMPS-Coll-msg.num}
	}
	\subfloat[Collective operation byte count.]{
		\includegraphics[width=.495\columnwidth]{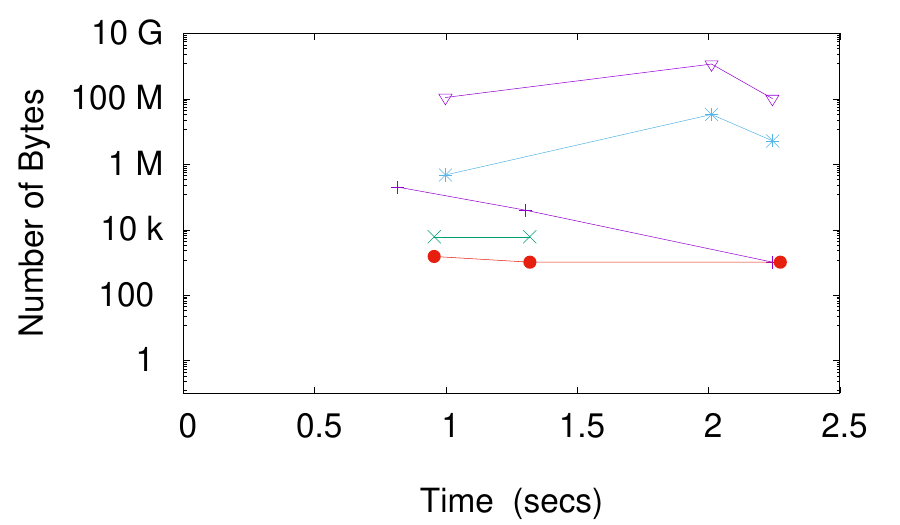}
		\label{fig:LAMMPS-Coll-bytes.num}
	}
 
 	\subfloat[Number of exchanged messages.]{
 		\includegraphics[width=.4\columnwidth]{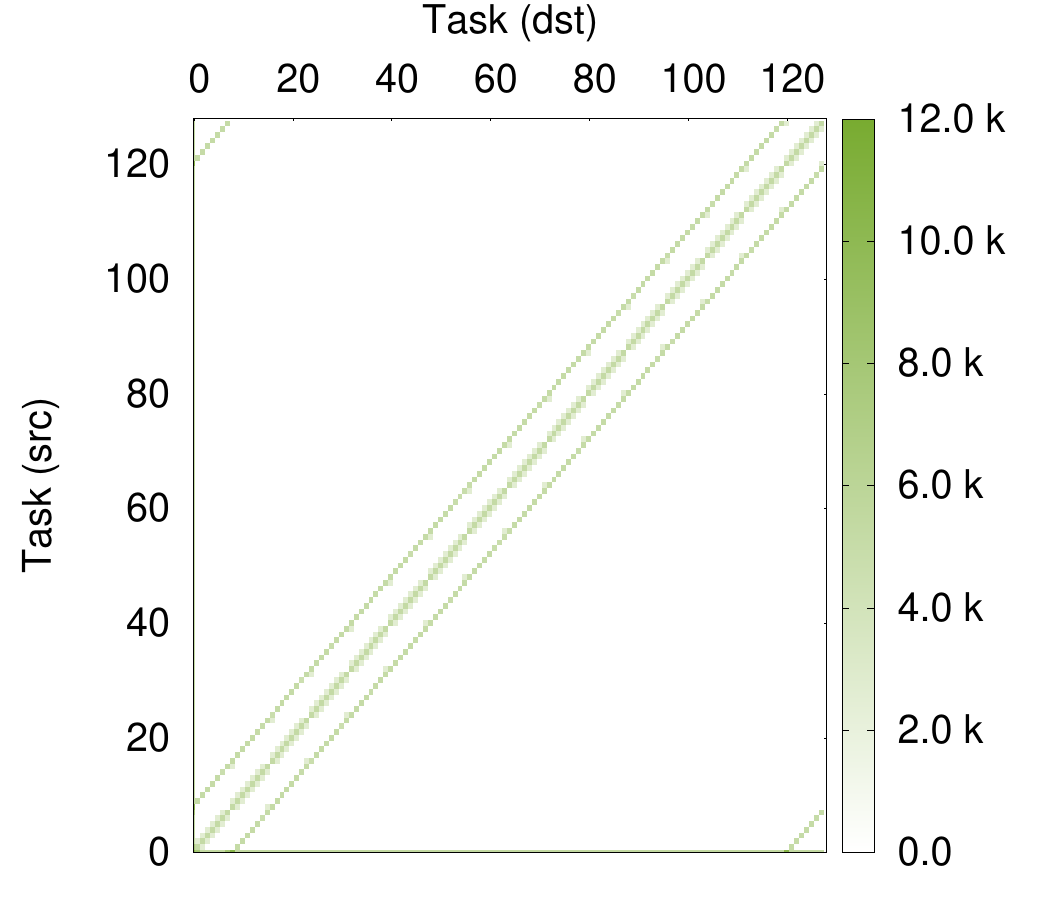}
 		\label{fig:LAMMPS-mat}
 	}
 	\subfloat[Number of exchanged bytes.]{
 		\includegraphics[width=.4\columnwidth]{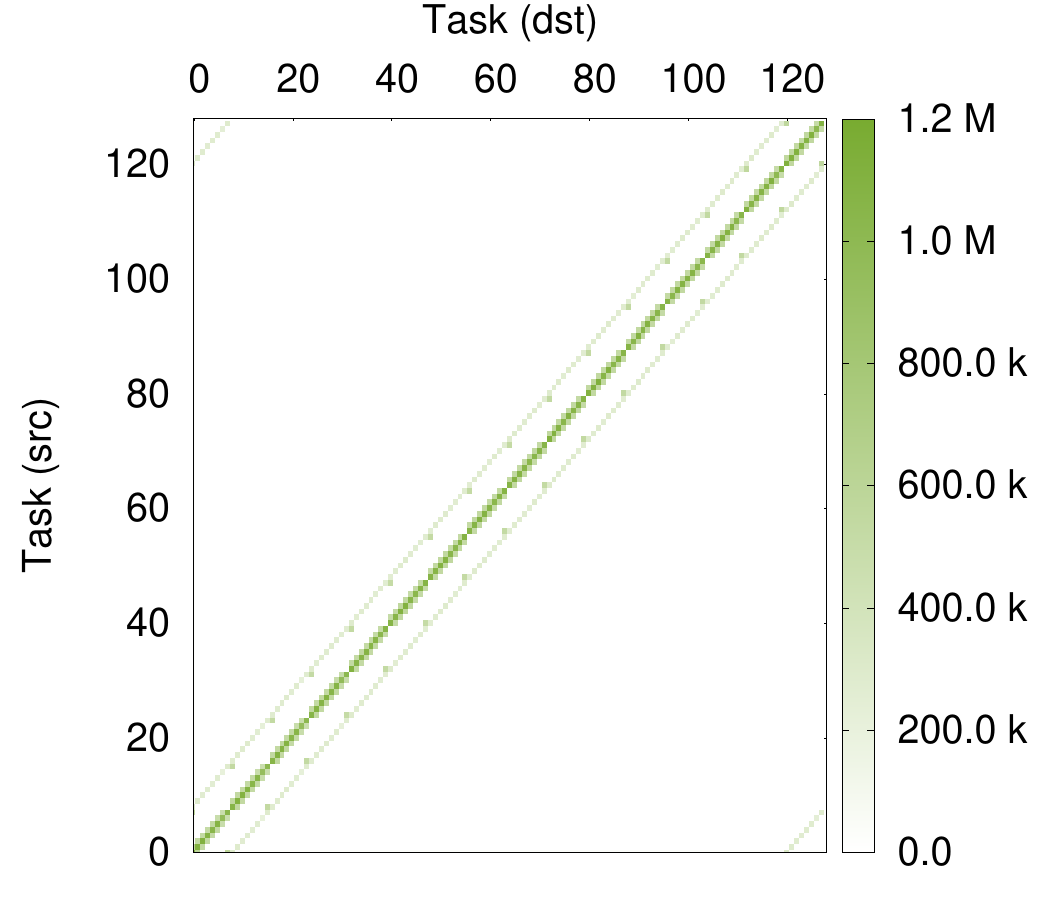}
 		\label{fig:LAMMPSB-bmat}
 	}
 	
 	\caption{LAMMPS static results.} 
 	\label{fig:LAMMPSstatic}
\end{figure}

 	

 
 	

Figure~\ref{fig:LAMMPSstatic} shows the static analysis results for the LAMMPS application.
Figures~\ref{fig:LAMMPS.num}~and~\ref{fig:LAMMPS.bnum} show the number of calls and bytes generated by the collective and point-to-point operations.
As we can see, \texttt{AllReduce} is the most predominant operation, followed by \texttt{P2P} and \texttt{Broadcast}.
Figures~\ref{fig:LAMMPS-Coll-msg.num}~and~\ref{fig:LAMMPS-Coll-bytes.num} show that LAMMPS starts broadcasting information among the nodes. Before one second passes, \texttt{Reduce} operations are called, and afterward, \texttt{AllReduce} operations are invoked.
After a while, more data is broadcast, and another reduction happens. At two seconds of runtime, a burst of \texttt{P2P} and \texttt{AllReduce} operations are called, and, finally, some hundreds of \texttt{Broadcast} and a million \texttt{AllReduce} messages mark the end of the execution.
Figure~\ref{fig:LAMMPS-Coll-bytes.num} shows that the generated bytes per collective are proportional to the collective call count.
Note that the communication pattern of LAMMPS, described in Section~\ref{sec:case-study:LAMMPS}, uses P2P operations to transfer data between two nearby domains, while \texttt{Broadcast} and, mainly, \texttt{AllReduce} collective operations are used to update the domain population.
Moreover, \texttt{AlltoAll} calls implement FFT operations required by long-range particle interactions.
Figures~\ref{fig:LAMMPS-mat}~and~\ref{fig:LAMMPSB-bmat} show that most traffic is either one-to-many or many-to-one, with MPI rank $0$ as the root. Note that \texttt{AllReduce}, while following a many-to-many communication pattern, behaves like a reduction and a broadcast with MPI task $0$ as the root rank. As with the NEST trace, most P2P messages are self-messages.

\begin{figure}[!htb]
	\centering
	\subfloat[Number of collective calls for each operation.]{
		\includegraphics[width=.495\columnwidth]{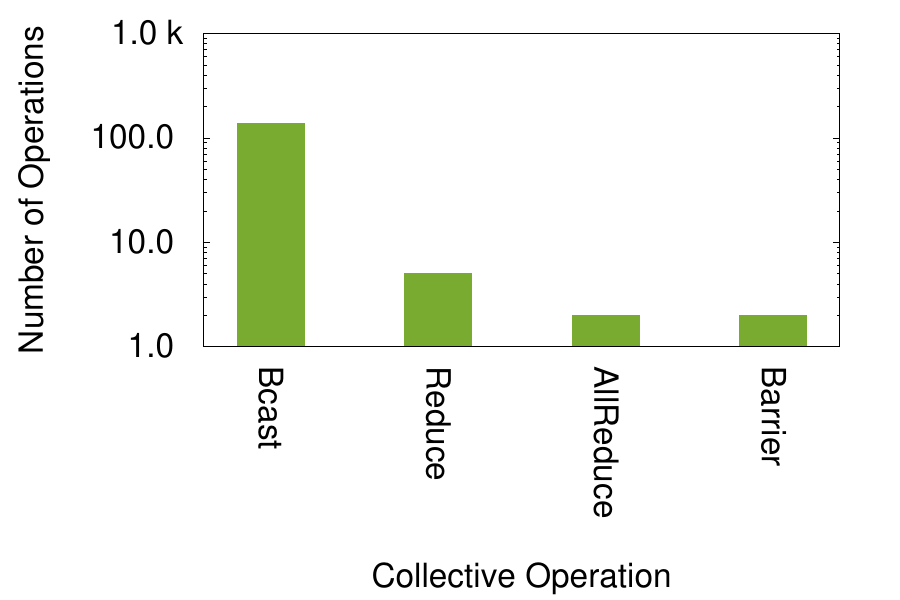}
		\label{fig:PATMOS.num}
	}
	\subfloat[Traffic generated by each operation.]{
		\includegraphics[width=.495\columnwidth]{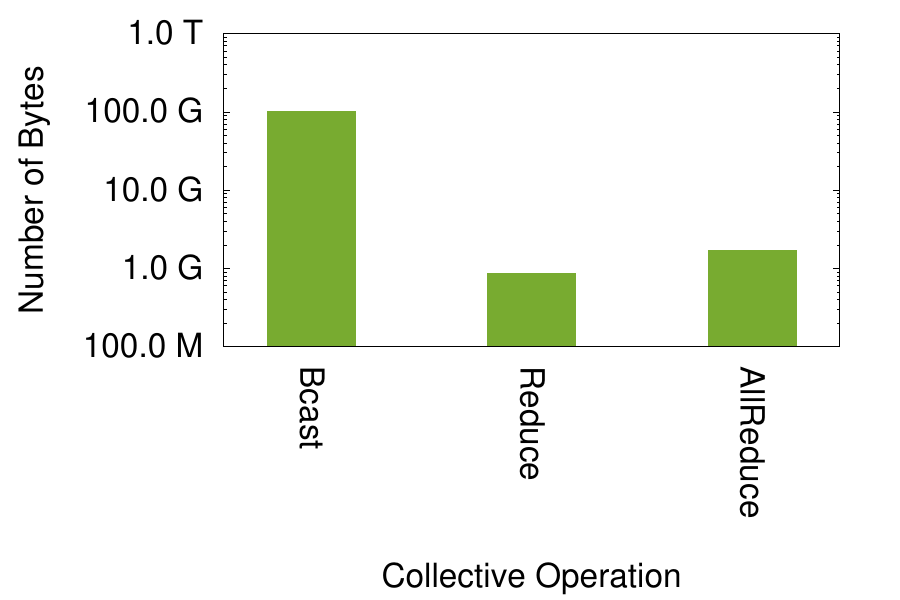}
		\label{fig:PATMOS.bnum}
	}

    \includegraphics[width=.75\columnwidth]{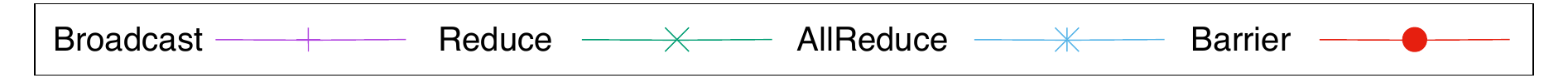}
    \subfloat[Collective operation message count.]{
		\includegraphics[width=.495\columnwidth]{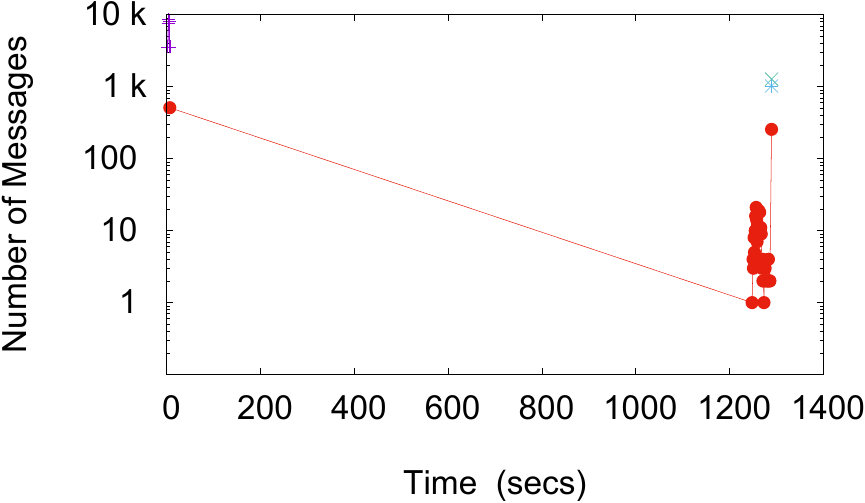}
		\label{fig:PATMOS-Coll-msg.num}
	}
	\subfloat[Collective operation byte count.]{
		\includegraphics[width=.495\columnwidth]{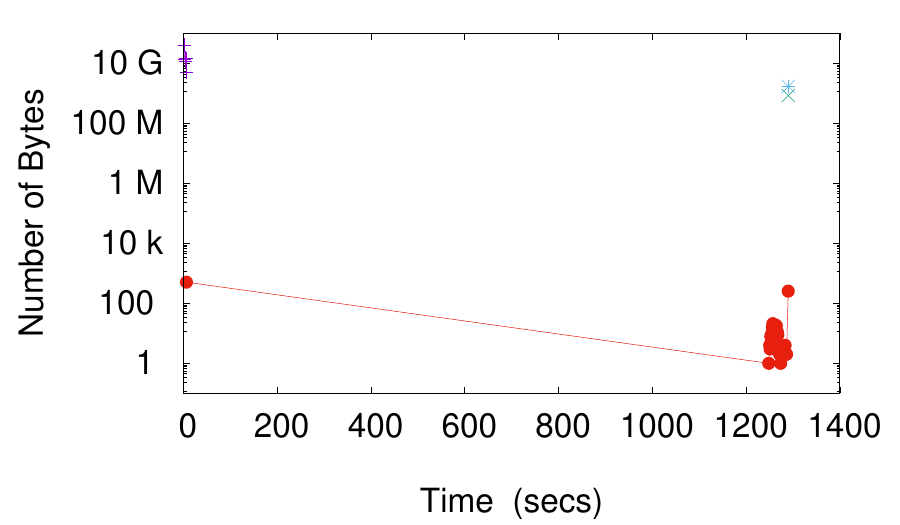}
		\label{fig:PATMOS-Coll-bytes.num}
	}
    \label{fig:PATMOSstatic2}
  
 	\subfloat[Number of exchanged messages.]{
 		\includegraphics[width=.4\columnwidth]{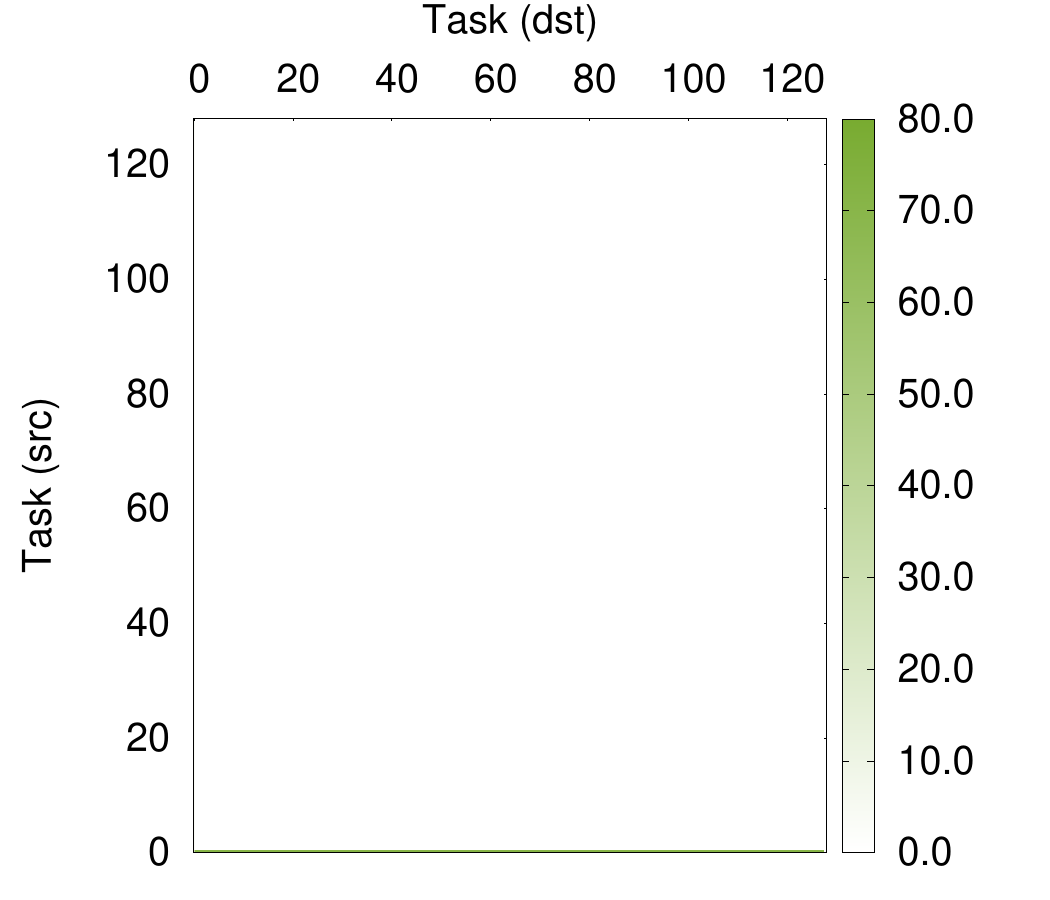}
 		\label{fig:PATMOS-mat}
 	}
 	\subfloat[Number of exchanged bytes.]{
 		\includegraphics[width=.4\columnwidth]{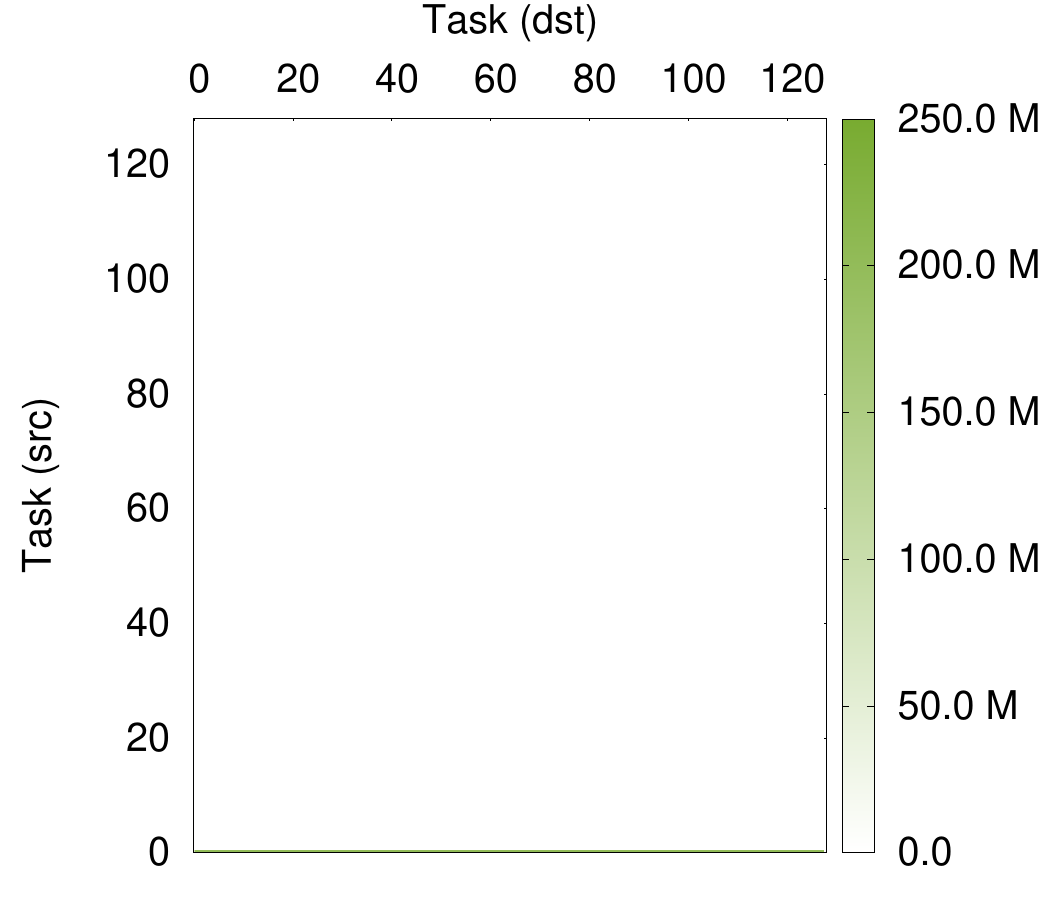}
 		\label{fig:PATMOSB-bmat}
 	}
 	\caption{PATMOS static results.} 
 	\label{fig:PATMOSstatic}
\end{figure}



  

Figure~\ref{fig:PATMOSstatic} shows the static analysis results for the PATMOS application.
Figures~\ref{fig:PATMOS.num}~and~\ref{fig:PATMOS.bnum} show that the most predominant operation is \texttt{Broadcast}, in terms of number of generated messages and bytes, followed by \texttt{AllReduce} and \texttt{Reduce}.
The communication of this application, as seen in Figure~\ref{fig:PATMOS-Coll-msg.num}, starts with two bursts of \texttt{Broadcast} with large amounts of data, after which nodes engage in computation using shared memory, so that no MPI calls are observed during a considerable period.
More than $1200$ seconds later, \texttt{AllReduce} and \texttt{Reduce} operations are performed.
Figure~\ref{fig:PATMOS-Coll-bytes.num} shows that the number of messages generated correlates with the number of generated bytes.
As described in Section~\ref{sec:case-study:PATMOS}, the computational model of PATMOS uses a hybrid parallel approach, where data is distributed among the processes that use multi-threading to operate on shared objects within each process.
These threads compute their contributions to the particle's history.
Only at the end of execution are the results of all processes shared among MPI ranks (i.e., via \texttt{AllReduce} and \texttt{Reduce} operations).
Figures~\ref{fig:PATMOS-mat}~and~\ref{fig:PATMOSB-bmat} show the messages and bytes exchanged by source and destination MPI ranks.
As we can see, the most common source/destination combination is from MPI rank 0 to the rest of the tasks, which means that most of the traffic is generated by the \texttt{Broadcast} operations.

\subsection{Dynamic analysis results}
\label{sec:ev:dynamic}

The \emph{dynamic analysis} evaluates network performance metrics when VEF traces are used to feed a network simulator, such as simulated runtime or FCT, and charts such as the cumulative distribution function (CDF) or buffering occupancy at network devices (see Section~\ref{sec:exp_conf}).
In this section, we analyze a set of metrics from the SAURON simulator, based on the results of the experiments described in Section~\ref{sec:exp_conf}.

\begin{figure}[!htb]
   \centering
   \subfloat[288-node Megafly.]{\includegraphics[width=0.37\textwidth, trim=0pt 0pt 0pt 20pt,clip]{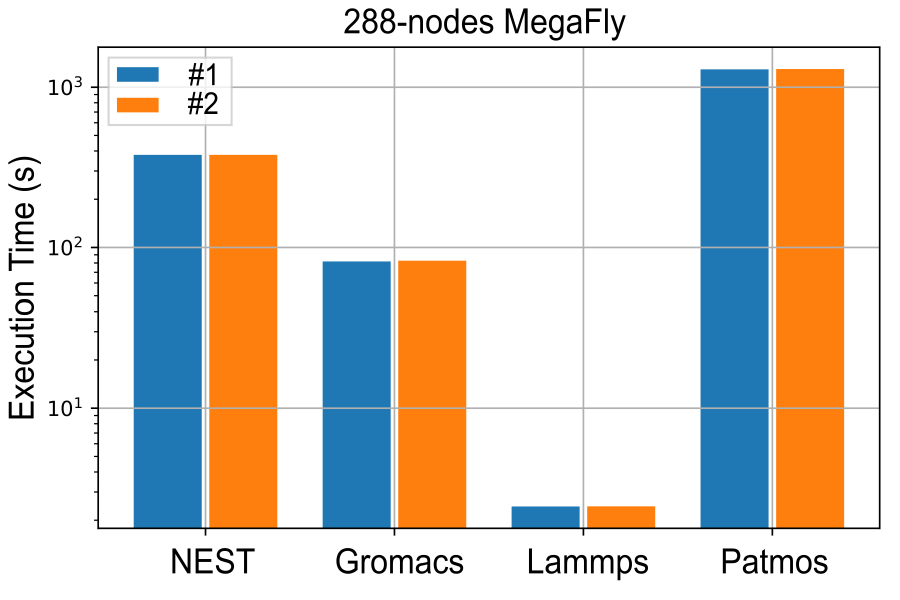}\label{fig:RunTimeFCT_DF+}}
   \subfloat[256-node Fat-Tree.]{\includegraphics[width=0.37\textwidth, trim=0pt 0pt 0pt 20pt,clip]{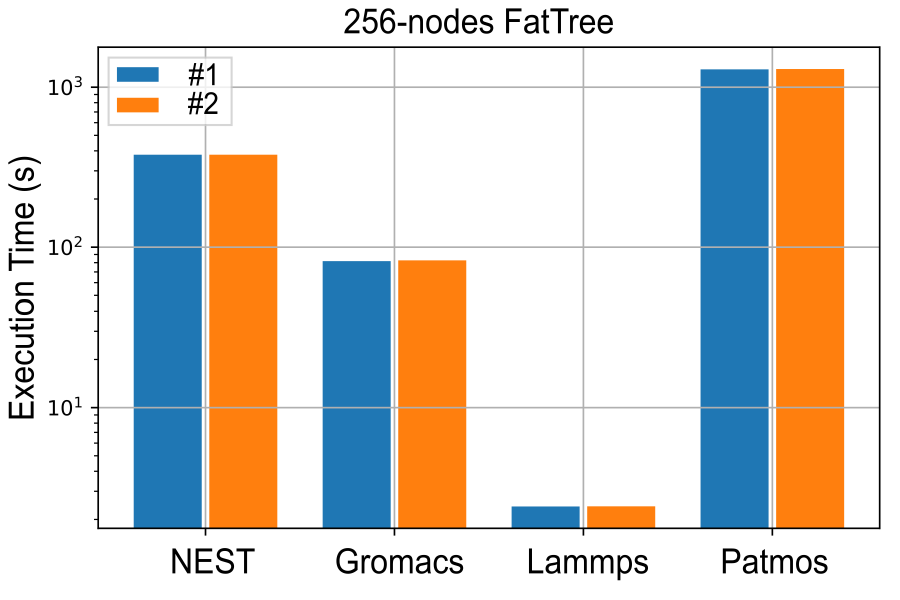}\label{fig:FCT_FT}}
   \caption{Execution time (seconds) for VEF traces of NEST, GROMACS, LAMMPS, and PATMOS, configured with 256 MPI ranks and comparing network configurations~$\#1$ and~$\#2$.}
   \label{fig:RunTimeTraces}
\end{figure}

Figure~\ref{fig:RunTimeTraces} shows the simulated execution time when we feed the SAURON simulator with VEF traces from the NEST, GROMACS, LAMMPS, and PATMOS applications. We have configured SAURON to model a 288-node Megafly and 256-node Fat-tree (see Table~\ref{tab:TopologiesCharacts}) and two network configurations defined in Table~\ref{tab:netconfs}. 

\begin{figure}[!htb]
   \centering
   \subfloat[288-node Megafly.]{\includegraphics[width=0.37\textwidth, trim=0pt 15pt 0pt 0pt,clip]{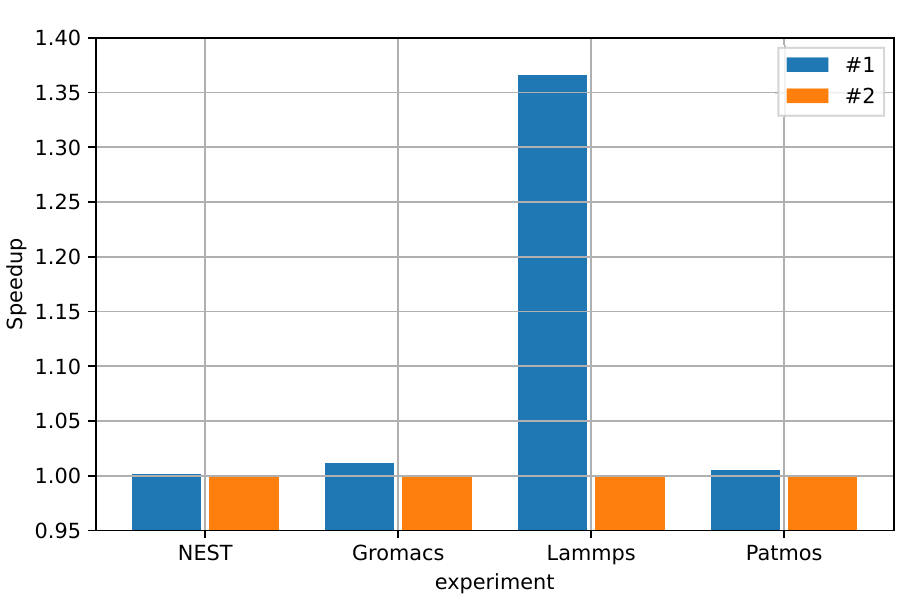}\label{fig:SpeedupDF+}}
   \subfloat[256-node Fat-Tree.]{\includegraphics[width=0.37\textwidth, trim=0pt 15pt 0pt 0pt,clip]{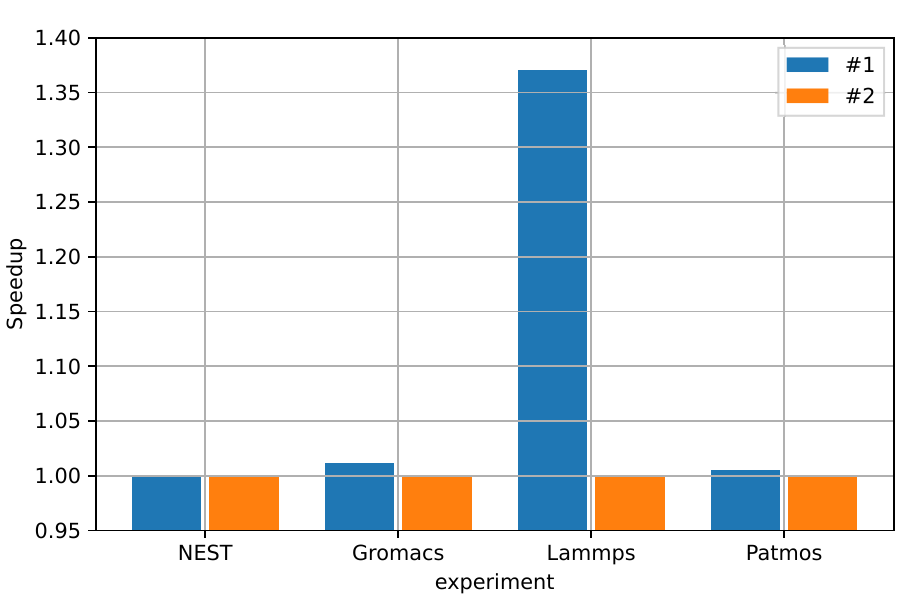}\label{fig:SpeedupFT}}
   \caption{Speedup obtained in configuration~$\#1$ compared against configuration~$\#2$.}
   \label{fig:SpeedupTraces}
\end{figure}

Note that the Execution time per application is the time the simulator takes to run completely the VEF trace, and that this is virtually identical to that observed in the static analysis (see Figures~\ref{fig:NESTstatic},~\ref{fig:GROMACSstatic},~\ref{fig:LAMMPSstatic},~and~\ref{fig:PATMOSstatic}). This confirms that the execution times measured in the dynamic analysis match those predicted by the static analysis. Since a large portion of application execution time is spent in idle wait rather than active network use, network configurations and topologies generally have little impact on execution time. But there are some exceptions, for instance, when simulating the LAMMPS application.
Figure \ref{fig:SpeedupTraces} illustrates the speedup achieved in execution time when comparing configuration~$\#1$ against configuration~$\#2$. 

As shown, there are negligible differences in execution time across network configurations and topologies because the traffic volume generated by these traces does not cause excessive network contention. LAMMPS is the only application in which a significant speedup is observed between the two configurations, reaching up to $1.35\times$. This occurs because it has the shortest execution time; consequently, the impact of the network architecture is more pronounced. Indeed, the time between messages recorded in the VEF trace (i.e., the time spent processing each MPI call) depends on the machine's CPU architecture.

\begin{figure}[!htb]
    \centering
    \subfloat[Mean FCT in a 288-node Megafly]{\includegraphics[width=0.4\textwidth]{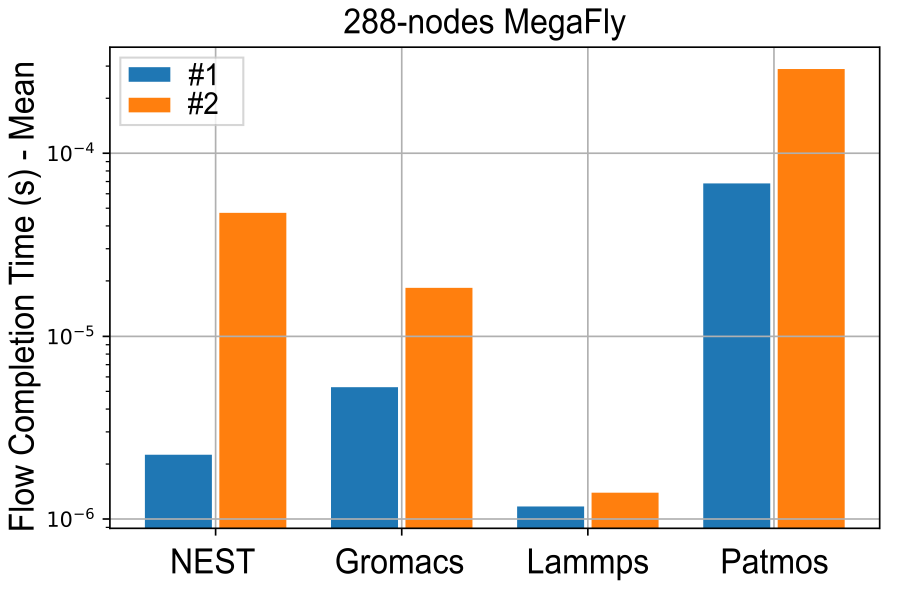}\label{fig:MeanFCT_DF+}}
    \subfloat[Mean FCT in a 256-node Fat-Tree]{\includegraphics[width=0.4\textwidth]{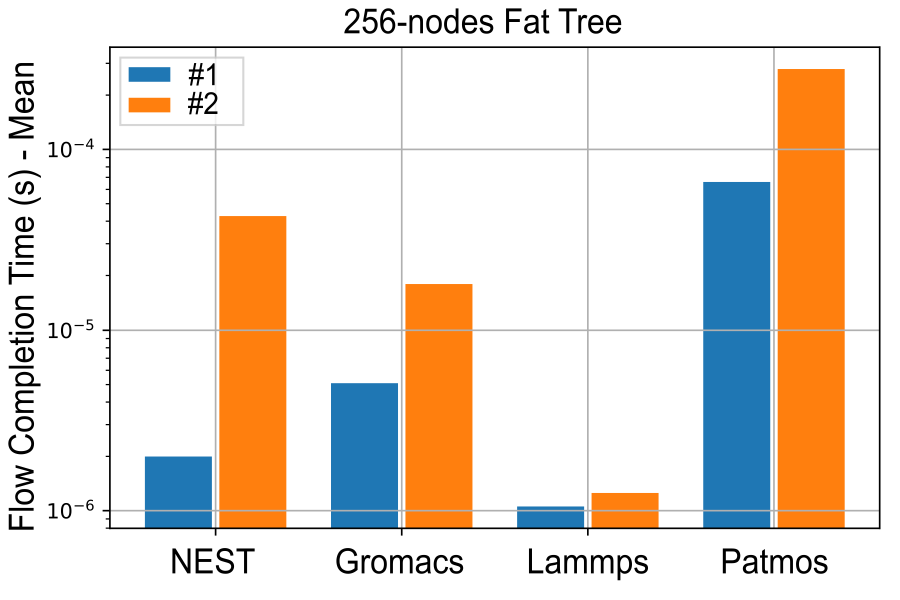}\label{fig:MeanFCT_FT}}
    \\
    \subfloat[Maximum FCT in a 288-node Megafly]{\includegraphics[width=0.4\textwidth]{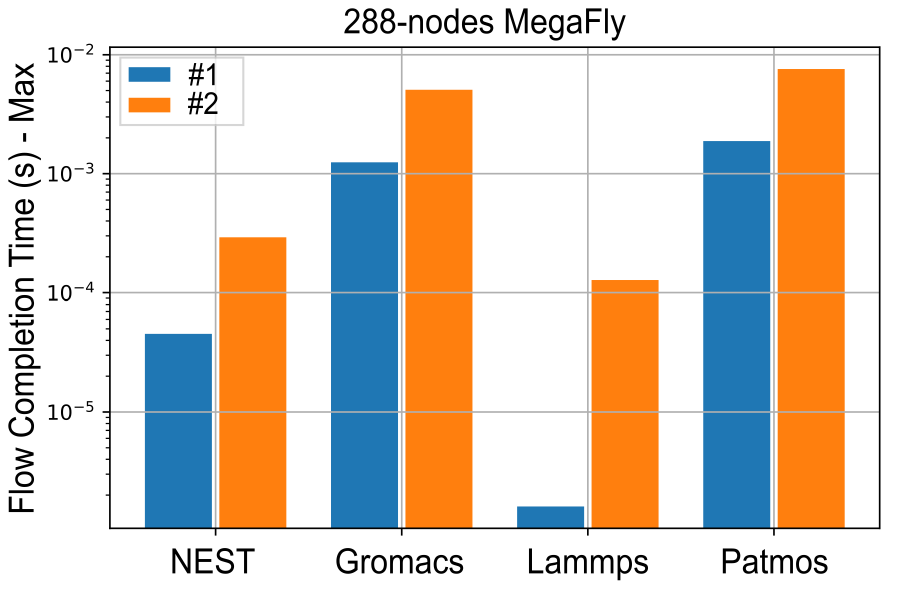}\label{fig:MaxFCT_DF+}}
    \subfloat[Maximum FCT in a 288-node Fat-Tree]{\includegraphics[width=0.4\textwidth]{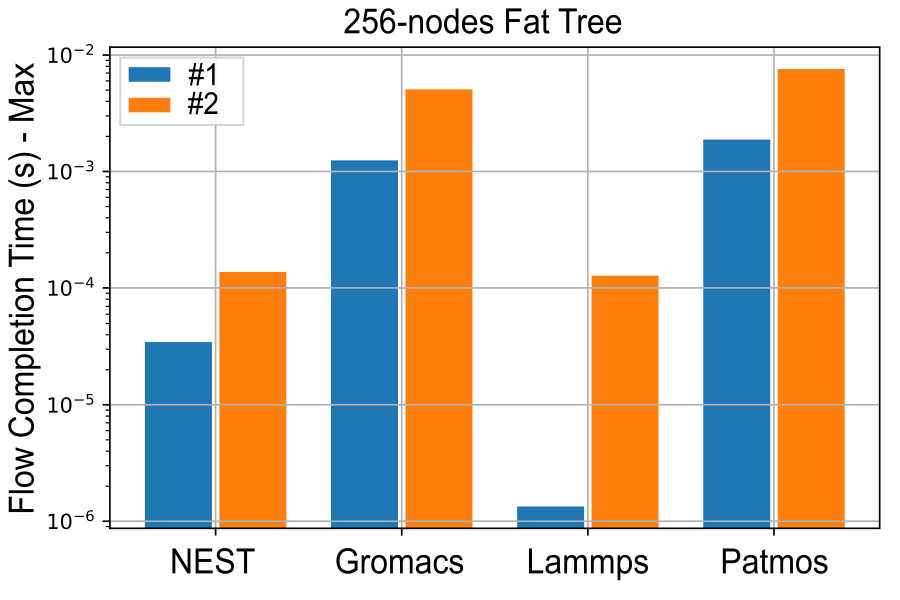}\label{fig:MaxFCT_FT}}
    \caption{Mean and Maximum FCT (in seconds) for VEF traces of NEST, GROMACS, LAMMPS, and PATMOS applications configured with 256 MPI ranks, comparing network configurations~$\#1$ and~$\#2$.}
    \label{fig:FCT}
\end{figure}

Figure~\ref{fig:FCT} illustrates the mean FCT for the same scenarios described above.
For every VEF trace and network topology, configuration~$\#1$ achieves a lower mean FCT because it is faster than configuration~$\#2$. It has a higher link data rate and MTU, and it uses variable packet size, allowing packets to be smaller than the MTU for short messages. In other words, the payload size for packets is only as long as required. Also, we note that for some applications, such as LAMMPS, the Fat-Tree topology achieves a slightly lower mean FCT than the Megafly results, since the former has a smaller diameter than the latter.

We observe a higher FCT value for the NEST application on the 288-node Megafly topology, due to the higher diameter. Despite having more switch ports available for distributing trace load in the 288-node Megafly topology, latency is nearly the same in both topologies. This means that processing time at the nodes is dominated by the computing times recorded in the trace file, which depend on the node architecture used to collect the specific VEF trace. However, we can appreciate that using Configuration~$\#1$ significantly reduces the maximum FCT compared to Configuration~$\#2$.

\begin{figure}[!htb]
	\centering
    \subfloat[NEST]{\includegraphics[width=0.45\textwidth]{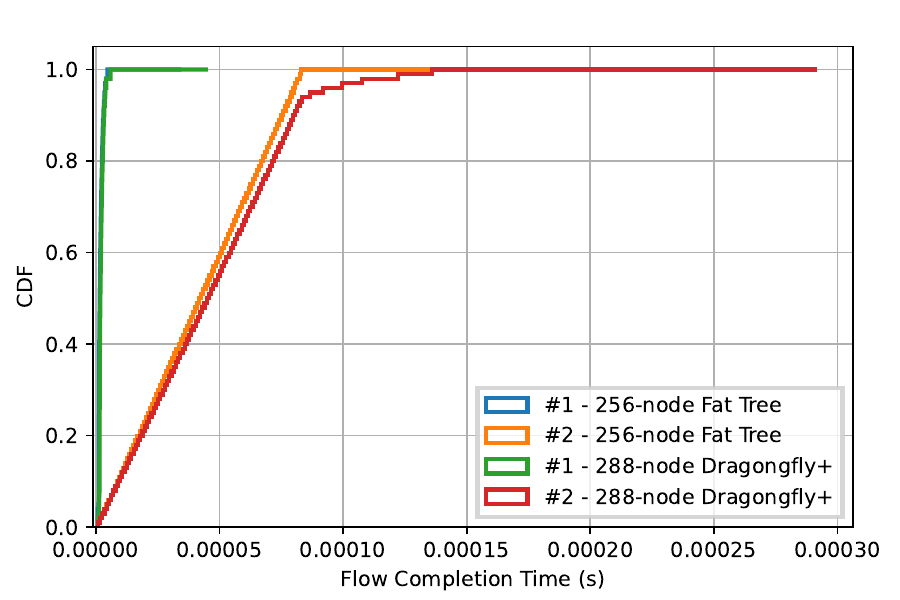}\label{fig:CDF_NEST}}
	\subfloat[GROMACS]{\includegraphics[width=0.45\textwidth]{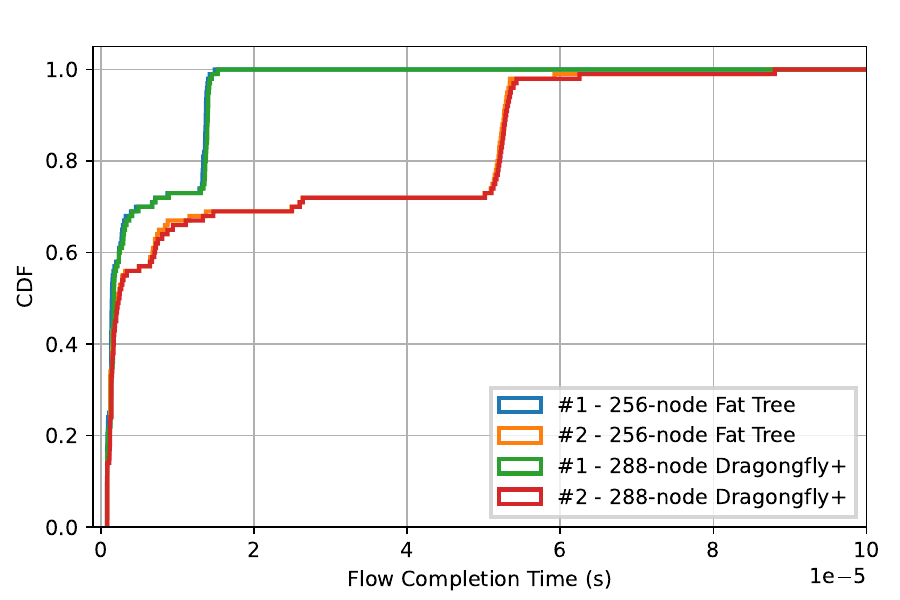}\label{fig:CDF_Gromacs}}\\ 
    \subfloat[LAMMPS]{\includegraphics[width=0.45\textwidth]{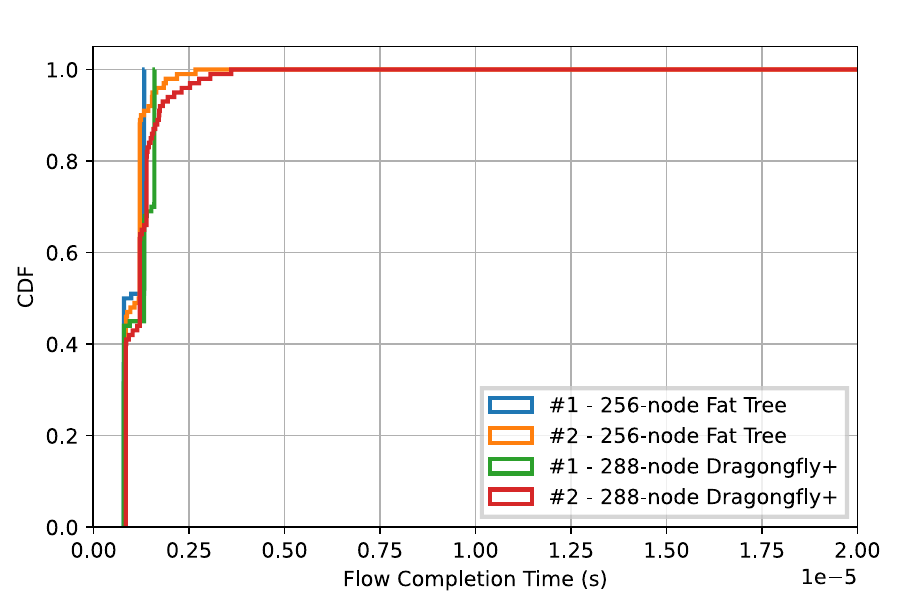}\label{fig:CDF_Lammps}}
    \subfloat[PATMOS]{\includegraphics[width=0.45\textwidth]{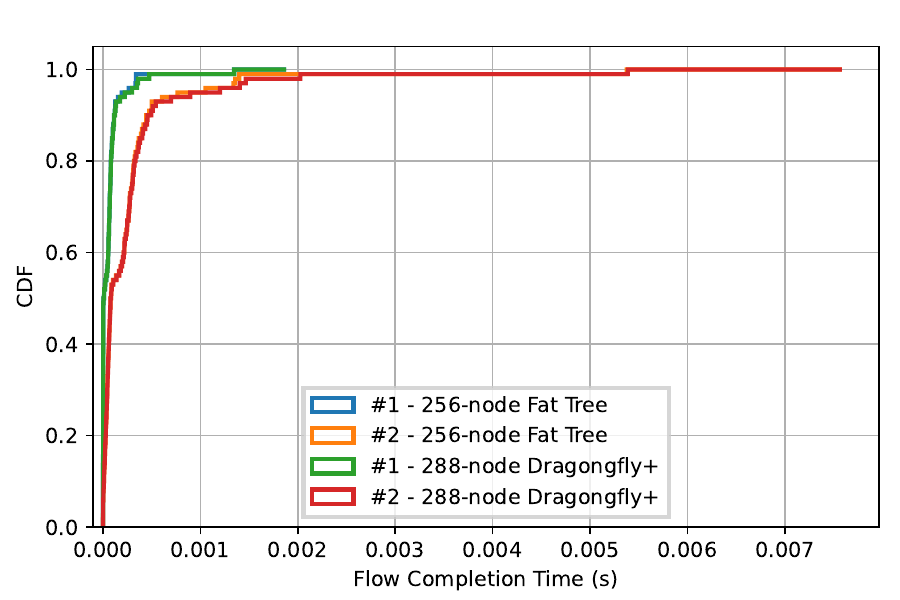}\label{fig:CDF_Patmos}}
	\caption{CDF of FCT for each trace, comparing topology and network architecture configuration.} 
	\label{fig:CDF_FCT}
\end{figure}

Figure~\ref{fig:CDF_FCT} shows the Cumulative Distribution Function (CDF) of the FCT metrics per application for each network configuration. Configuration~$\#1$ obtains better FCT values than Configuration~$\#2$, regardless of the network topology. Moreover, the Megafly topology yields slightly worse results than the Fat Tree.
For the NEST application (see Figure~\ref{fig:CDF_NEST}), Configuration~$\#1$ obtains significantly better FCT values than Configuration~$\#2$.
For Conf.~$\#1$, 100\% of flows finish in less than $5$ms.
For GROMACS (see Figure~\ref{fig:CDF_Gromacs}), similar conclusions can be drawn, with a special focus on the step produced in the FCT series after 70\% of the flow completion, where latency moderately increases for  Conf.~$\#1$ while significantly increases for Conf.~$\#2$, since GROMACS generates a moderate load that delays the completion of medium-sized and large flows.
Similar to NEST, we can see for LAMMPS (see Figure~\ref{fig:CDF_Lammps}) and PATMOS (see Figure~\ref{fig:CDF_Patmos}) applications that Conf.~$\#1$ completes 100\% of its traffic flows with significantly lower FCT values than those obtained by Conf.~$\#2$, and the tail latency of the latter is by far the largest one.

To extend the dynamic analysis, we must examine additional FCT metrics to identify why tail latency increases significantly in specific network scenarios.
The following section extends our study by characterizing congestion episodes in the applications we analyzed.

\subsection{Congestion characterization}

\begin{figure}[!htb]
	\centering
    \subfloat[Fat Tree (Conf.~\#1).]{\includegraphics[width=0.499\textwidth, trim=45pt 95pt 30pt 212pt,clip]{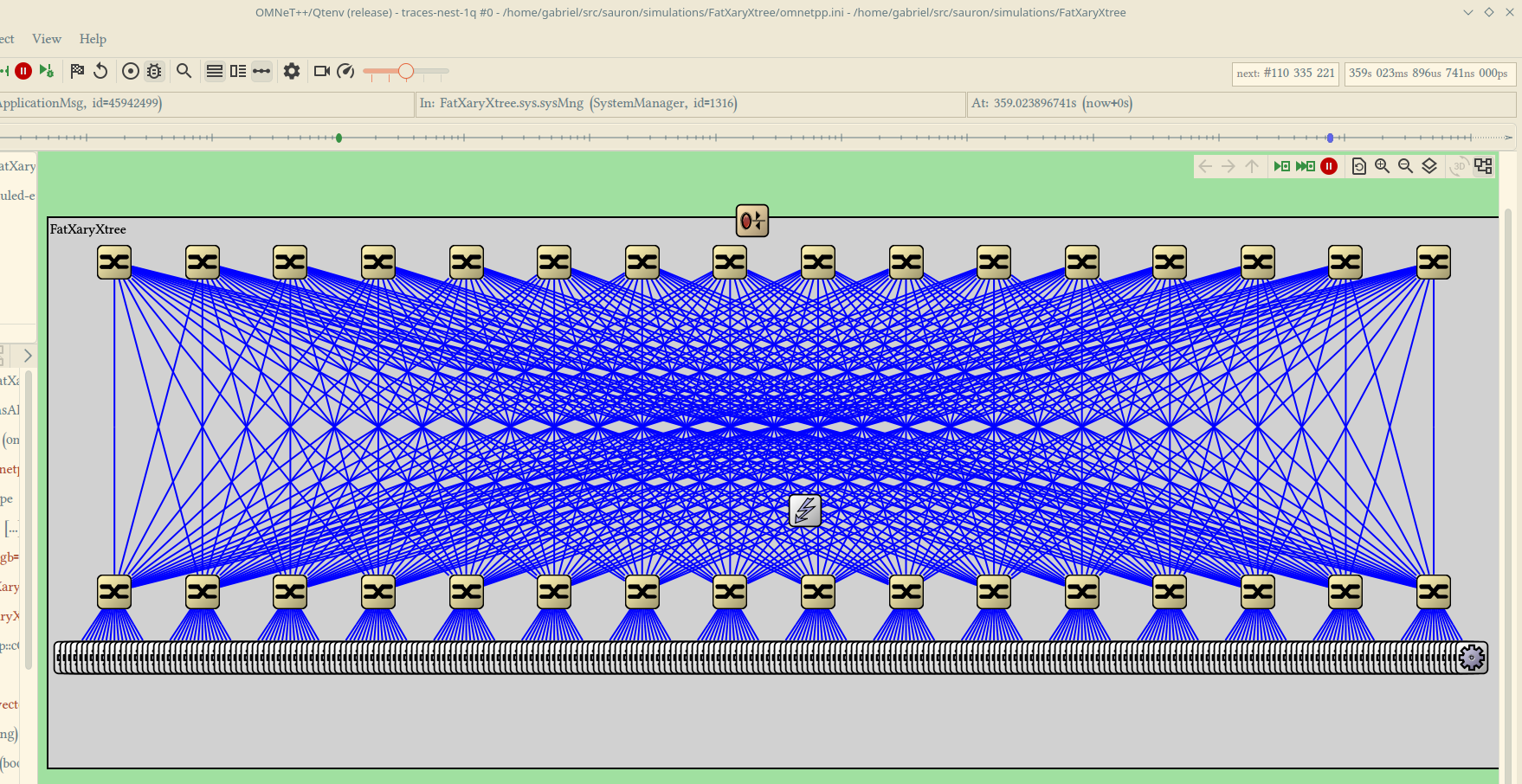}\label{fig:occpNestFtree1}}\hfil
	\subfloat[Fat Tree (Conf.~\#2).]{\includegraphics[width=0.499\textwidth, trim=20pt 88pt 10pt 185pt,clip]{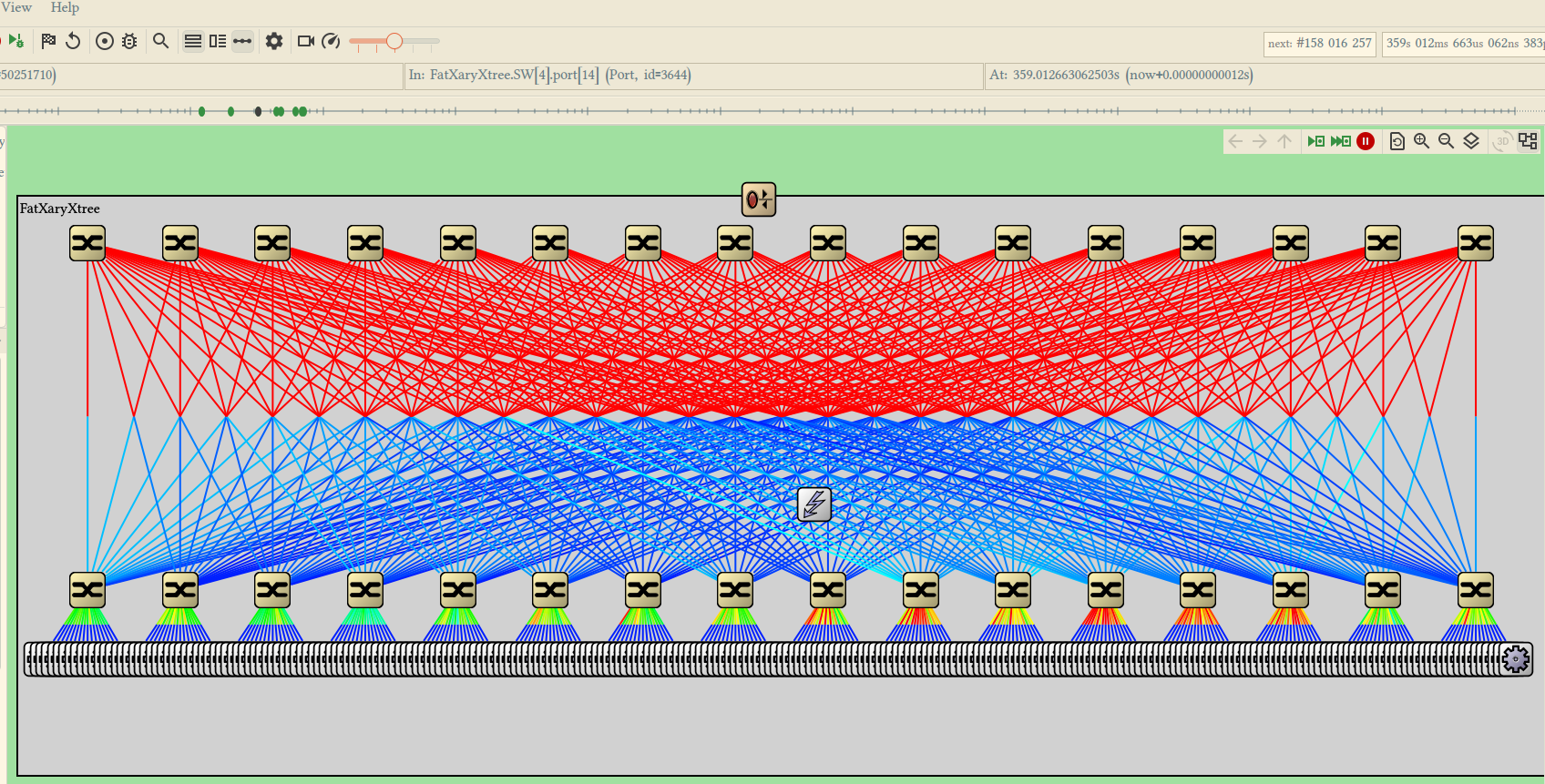}\label{fig:occpNestFtree2}}\\
    \subfloat[Megafly (Conf.~\#1).]{\includegraphics[width=0.49\textwidth, trim=88pt 10pt 12pt 145pt,clip]{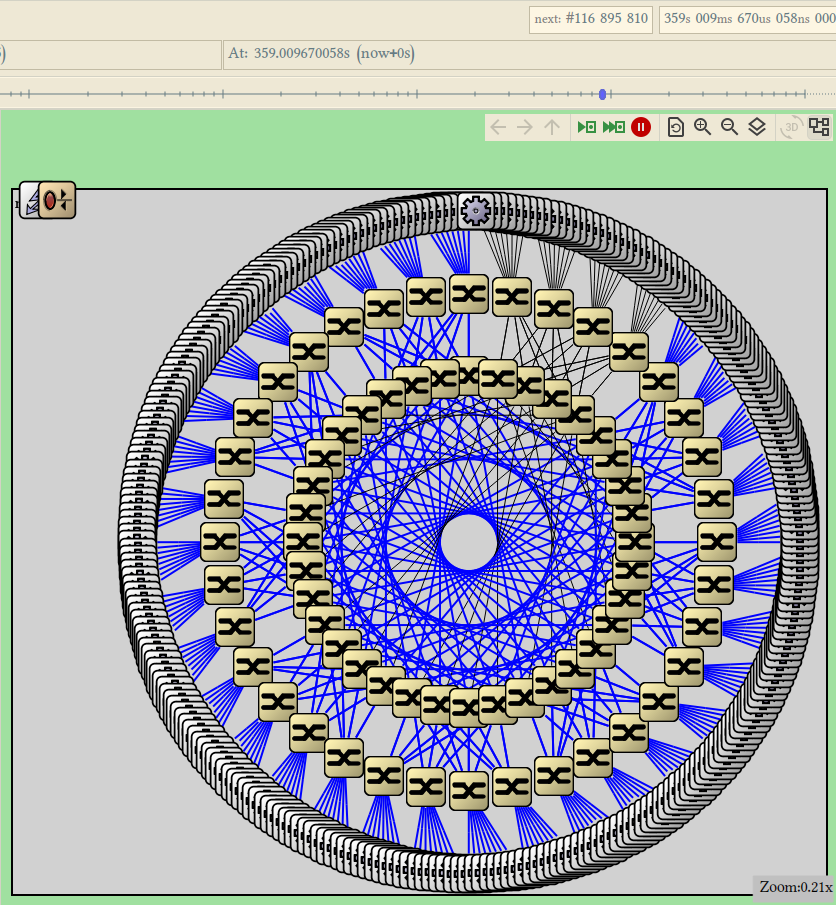}\label{fig:occpNestMFly1}}\hfill
	\subfloat[Megafly (Conf.~\#2).]{\includegraphics[width=0.49\textwidth, trim=88pt 10pt 17pt 145pt,clip]{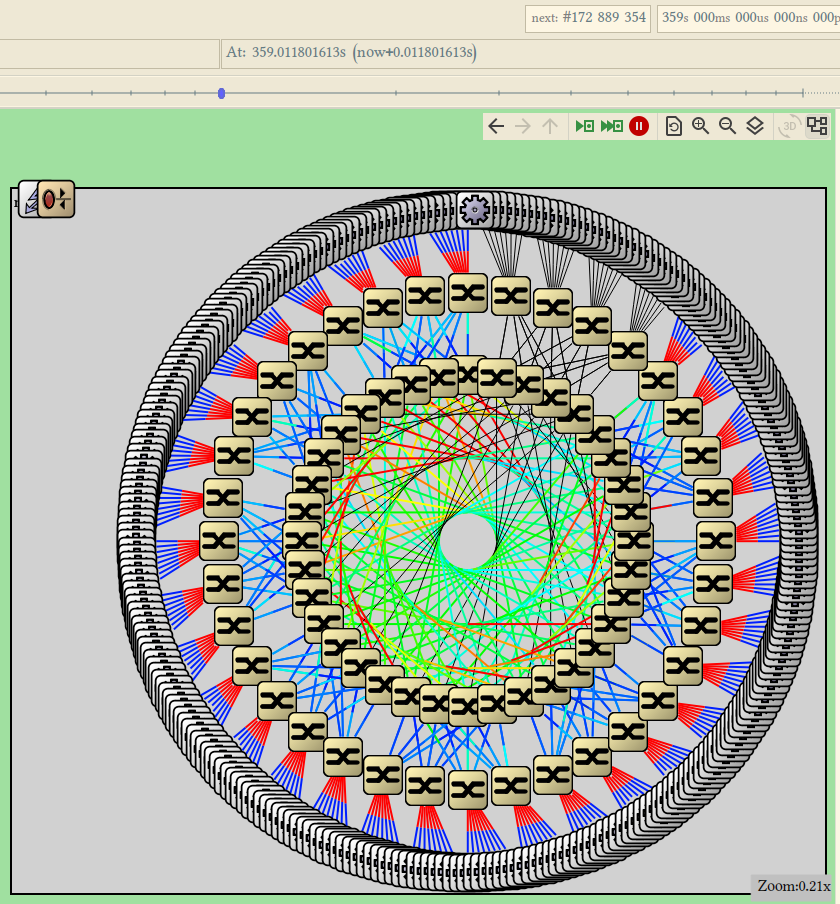}\label{fig:occpNestMFly2}}
    \caption{Maximum input occupation in NEST from 358.9 to 359 seconds.}
    \label{fig:OccupancyNest}
\end{figure}

An indicator of the severity of network congestion is the occupancy of the queues at switches’ ports.
In this section, we analyze the occupation of input ports in switches using the new GUI in the SAURON simulator with VEF traces.
This GUI depicts the topology's switches and links in a smooth layout, making it easy to identify hotspots and congestion dynamics induced by the traffic patterns at a given time instant.
The input buffer occupancy is shown by coloring half of the links; a link connected to a switch or NIC is colored based on the occupancy of the input port buffer for that switch or NIC.
The logic behind the link coloring is as follows: green indicates empty buffers, blue indicates low occupancy, and red indicates the highest occupancy level in a period of time. Any color displayed between those is calculated using a gradient for higher granularity. This way, we can visualize the presence of congestion trees and their potential backpropagation across the network.
Figure~\ref{fig:OccupancyNest} shows the maximum input occupation of the Fat-tree and Megafly topologies (see Table~\ref{tab:TopologiesCharacts}) and network architectures \#1 and \#2 (see Table~\ref{tab:netconfs}) for the NEST application.

According to the static analysis (see Figure~\ref{fig:NESTstatic}), after 358 seconds of execution time, the communication is dominated by \texttt{AlltoAll} collective operations, producing a moderate load. For network architecture \#1, we can see that links are mostly colored in blue, meaning that there is traffic exchange with low buffer occupancy (see Figures~\ref{fig:occpNestFtree1} and~\ref{fig:occpNestMFly1}). Note that we use 400~Gbps links in this configuration. However, in network architecture \#2, i.e., small buffering and lower link bandwidth (100~Gbps), we can see that occupancy is high in the Fat Tree (see Figure~\ref{fig:occpNestFtree2}), at the input buffers of the switches at the second stage. We can also see end nodes that are visited more than others, i.e., those with links colored in red, indicating that congestion originates at those nodes.
By contrast, in the Megafly topology and network architecture \#2 (see Figure~\ref{fig:occpNestMFly2}), we observe congestion in the input buffers of the first-level switches and in some inter-group communications.

\begin{figure}[!htb]
	\centering
    \subfloat[Fat Tree (Conf.~\#1).]{\includegraphics[width=0.499\textwidth, height=2.5cm, keepaspectratio=false]{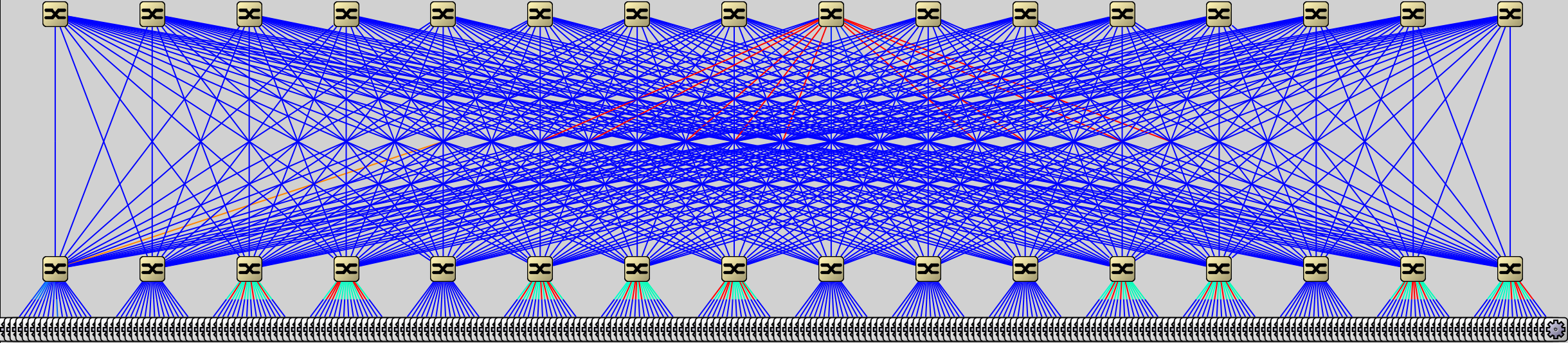}\label{fig:occpNestIncastFtree1}}\hfill
	\subfloat[Fat Tree (Conf.~\#2).]{\includegraphics[width=0.499\textwidth, height=2.5cm, keepaspectratio=false]{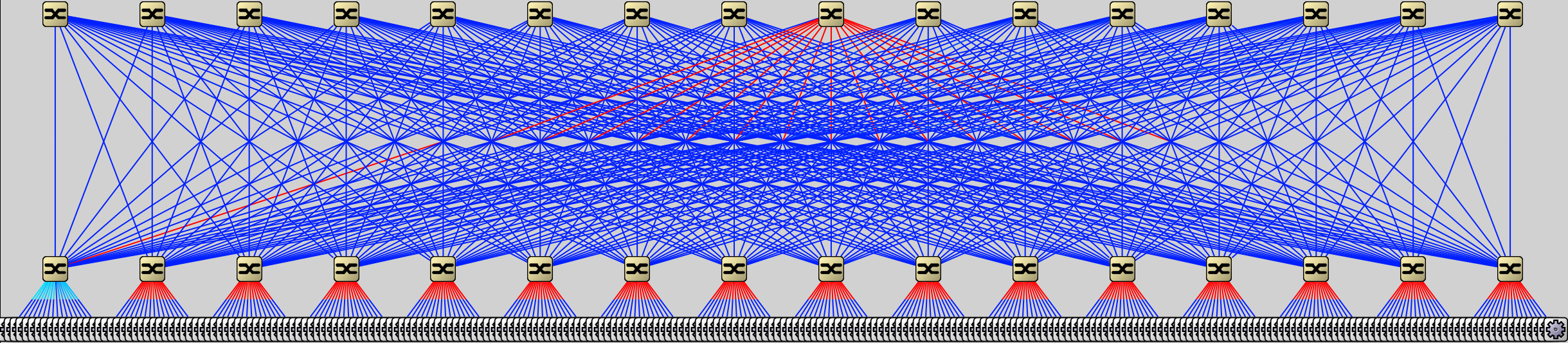}\label{fig:occpNestIncastFtree2}}\\
    \subfloat[Megafly (Conf.~\#1).]{\includegraphics[width=0.49\textwidth]{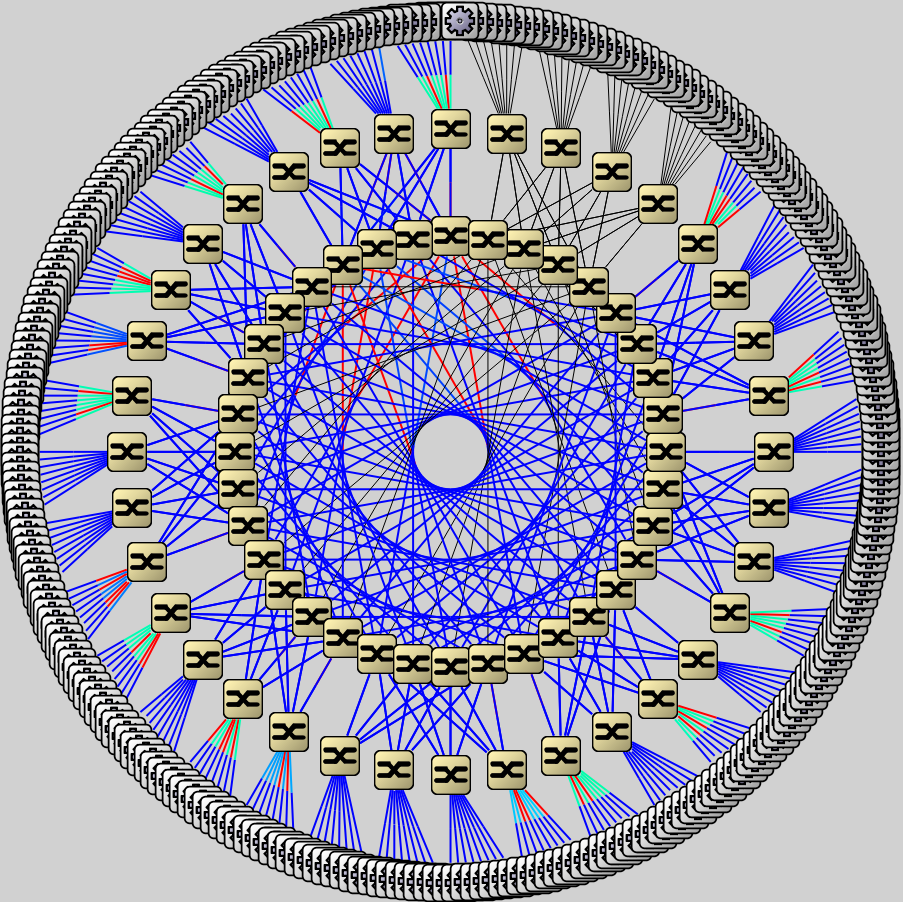}\label{fig:occpNestIncastMFly1}}\hfill
	\subfloat[Megafly (Conf.~\#2).]{\includegraphics[width=0.49\textwidth]{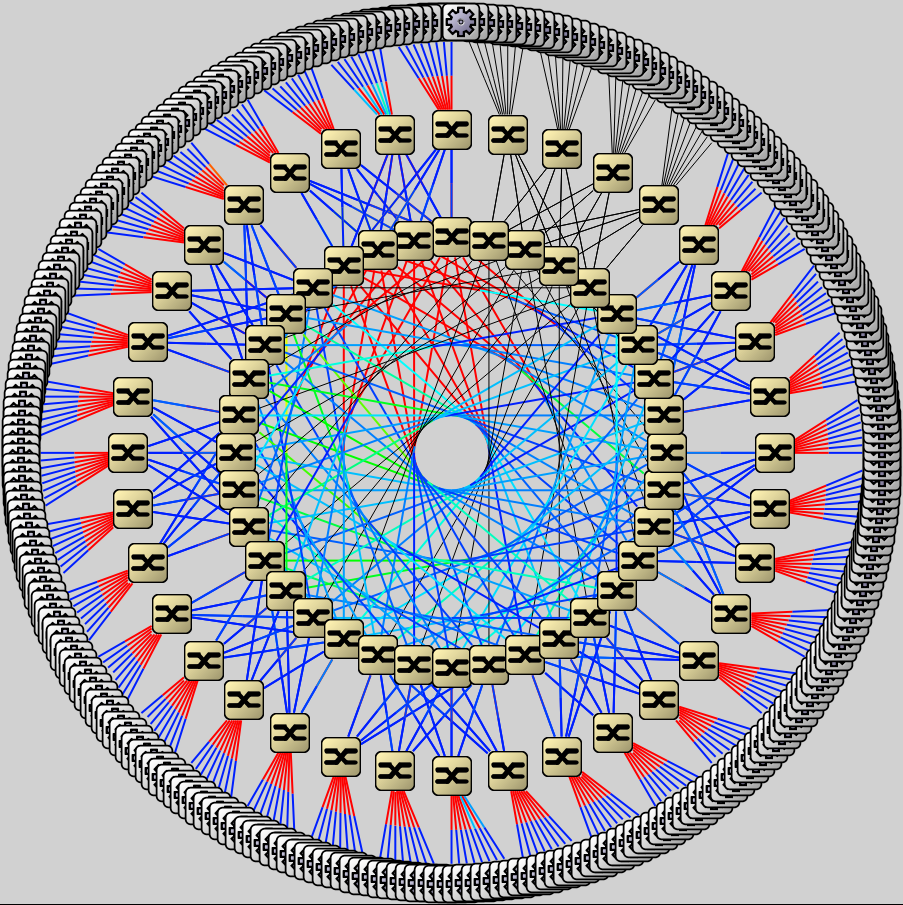}\label{fig:occpNestIncastMFly2}}
    \caption{Maximum input occupation in NEST from 358.9 to 359 seconds, when the incast traffic is injected.}
    \label{fig:OccupancyNest+incast}
\end{figure}

\begin{figure}[!htb]
    \centering
    \subfloat[Mean Flow Completion Time]{\includegraphics[width=0.8\textwidth]{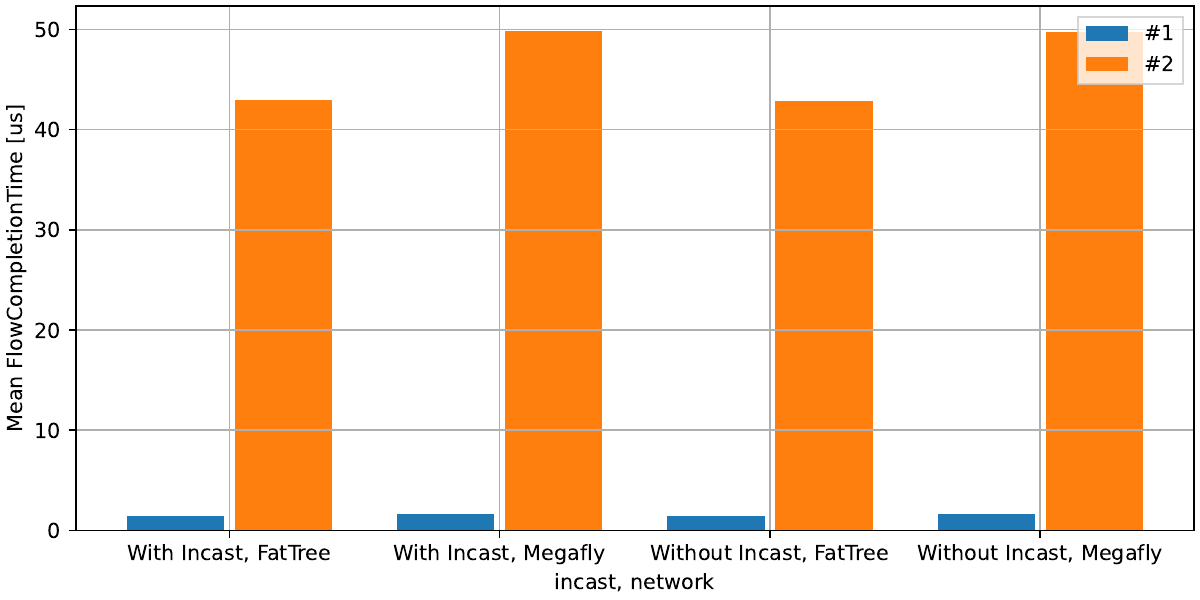}\label{fig:MeanFCTNestIncast_DF+}}
    \\
    \subfloat[Maximun Flow Completion Time]{\includegraphics[width=0.8\textwidth]{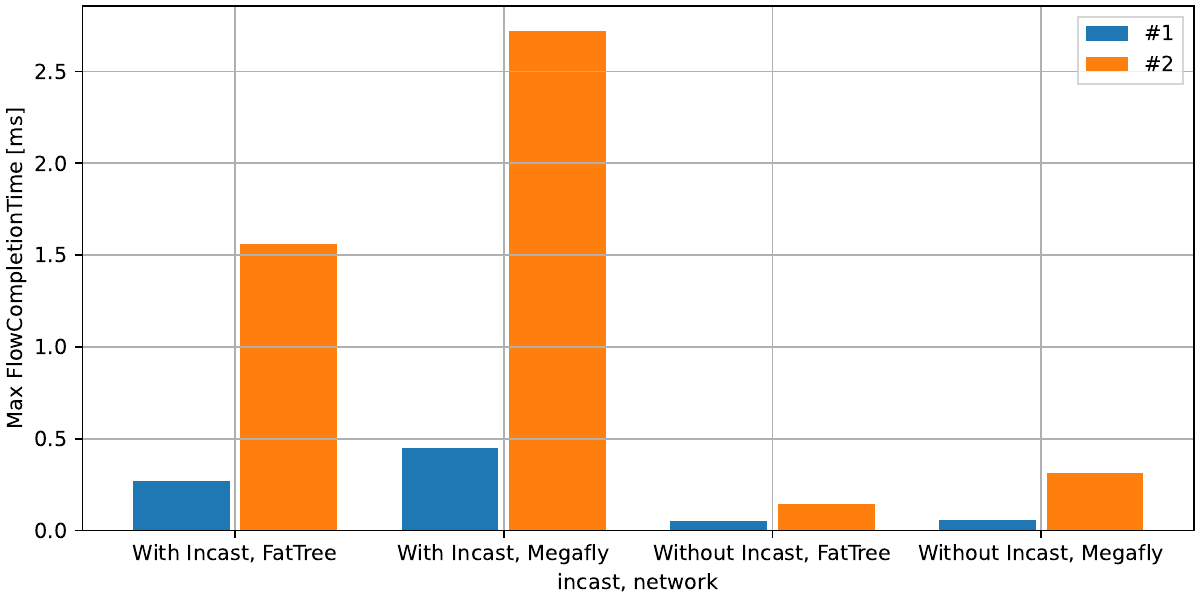}\label{fig:MaxFCTNestincast_DF+}}
    
    \caption{Mean and Maximum FCT for NEST, configured with 256 MPI ranks and comparing network configurations~$\#1$ and~$\#2$, in both topologies, and with and without incast.}
    \label{fig:Nest+Incast_FCT}
\end{figure}

To analyze the impact of congestion in greater detail, we combined the previously analyzed Nest trace with a synthetic traffic burst. In this scenario, $64$ nodes send a \qty{10}{\mebi\byte} message to a single destination node simultaneously. Figure~\ref{fig:OccupancyNest+incast} shows the maximum input port occupancy during this interval. As observed, this burst has a more significant impact on configuration~$\#2$ across both topologies. It is also notable that the maximum occupancy in the remaining ports is considerably lower. This suggests that the congestion tree generated by incast traffic reaches the source nodes, thereby negatively impacting overall network performance. This observation is corroborated by Figure~\ref{fig:Nest+Incast_FCT}, which shows that although the mean FCT is unaffected by incast traffic, the maximum FCT increases substantially across both topologies and configurations. The highest maximum FCTs are observed in the Megafly topology with configuration~$\#2$.

We provide congestion characterization for the remaining traces in the supplementary PDF. 

\section{Related Work}
\label{sec:related-work}

Reproducing the traffic patterns of real workloads in simulators is vital for comparing our simulation tools against real systems. Several tools exist to model and characterize traffic, such as  Vampir~\cite{vampir08}, Score-P~\cite{scorep12}, Scalasca~\cite{scalasca10}, DUMPI~\cite{sst-dumpi}, or Extrae~\cite{extrae-doc}. However, these tools may not be suitable for network simulations due to trace size or the use of absolute timestamps. For instance, traces that contain more information than necessary (e.g., message source, destination, and size) may be slower or more challenging to process. Moreover, using absolute timestamps means that time-related information is tightly tied to the system and network on which traces were collected, and this is reflected in the traffic patterns. To address this, Andujar et al. proposed VEF self-related traces~\cite{Andujar15cluster}, which include only information used in network simulators. Relative timestamps and self-related messages help extract application traffic patterns, leaving network behavior to the simulator. The framework consists of VEF-Prospector~\cite{VEF-Prospector}, a profiling program that generates MPI calls for running applications, and VEF-TraceLib~\cite{VEF-TraceLIB}, a library for use in simulators. VEF-TraceLib can also synthesize patterns from other traffic models~\cite{Montazeri18sigcomm}. After capturing MPI calls, VEF-Prospector also includes a static analysis tool, as shown in this paper.

On the other hand, interconnection network simulators provide designers with insights into expected network performance that would otherwise be difficult to obtain using other methodologies or tools. 
Several network simulators are available. 
TOPAZ~\cite{Abad12topaz} is a general-purpose interconnection network simulator that can model a lot of different switches and gives the functionality to sacrifice precision for the sake of speed.
NS-3~\cite{Riley10ns3} is a discrete-event network simulator for Internet systems, targeted primarily for research and educational use.
It is primarily based on NS-2, which was initially devised to model TCP. Ns-3 was designed from the ground up to convey higher detail and realism, bringing actual software implementations closer to the
real-world models.
The Structural Simulation Toolkit (SST)~\cite{Rodrigues11sst} is a modular simulation framework for modelingdesigned to model high-performance computing systems at multiple levels of abstraction. It facilitates hardware and software co-design by enabling detailed simulations of processors, memory hierarchies, and interconnects. SST supports parallel execution and scalability, allowing researchers to explore large-scale architectural configurations. Its component-based architecture promotes extensibility and integration with third-party tools. SST is widely used in architectural research for performance prediction and design-space exploration.
OMNeT++~\cite{Omnetpp} is a general simulation environment designed to provide a simulation framework for developers and researchers in distributed systems, computer networks, and multiprocessors. Specific simulation models can be built in OMNeT++ to simulate domains such as wireless networks, business processes, P2P networks, or high-performance interconnection networks. Some well-known OMNeT-based models include INET~\cite{vejrazka2013inet} and the SAURON simulator~\cite{Sauron}, which were used in this study.
The SAURON simulator has recently been integrated with other simulators, such as COSSIM, which models the system-node architecture with higher accuracy \cite{TampouratzisPGREG25}, enabling us to generate VEF traces at different message-generation rates. In this paper, we have assumed constant message-generation rates for the systems used to collect VEF traces.
Recently, other simulation frameworks, such as Astra-Sim~\cite{won2023astrasim2} or ATLAHS \cite{shen2025atlahs}, enable systematic investigation of modern HPC and deep learning workloads, supporting the identification of bottlenecks and the design of efficient methodologies for large-scale DNN models across diverse future platforms. Through its APIs, users can seamlessly integrate any custom-made network, compute, or memory simulator backend in a plug-and-play manner. Integrating the methodology proposed in this paper with the Astra-Sim or ATLAHS APIs is left for future work. Additionally, the DUMPI, GOAL, and Chakra trace-file formats used in SST, ATLAHS, and Astra-sim, respectively, should be considered and compared with the VEF format to discuss the advantages and disadvantages of these three formats.

\section{Conclusions}
\label{sec:conclusions}

In this paper, we present an open-source framework and methodology for characterizing communication patterns from real MPI-based applications running on HPC systems and converting these patterns into specific-purpose VEF traces, enabling network designers and researchers to use them to feed network simulators.
We have analyzed a set of VEF traces from a public repository, which were collected through collaboration across several EU-funded projects (EuroHPC-JU).
These traces were obtained from real runs of NEST, GROMACS, LAMMPS, and
PATMOS applications in different HPC systems.
To analyze the behavior of these traces, we performed static analysis to measure key metrics, such as the amount of data generated by communication operations, which may indicate network congestion.
We also performed dynamic analysis of network performance metrics using the OMNeT++-based SAURON interconnection network simulator. The analysis results provide insight into congestion generated in the network by the mentioned applications.
The traces repository and the open-source framework are available to the community to either populate that repository with VEF traces from new application runs or use the available ones in their network simulators.

\section*{Acknowledgment}

This work has been supported by the RED-SEA project, which received funding from the European High-Performance Computing Joint Undertaking (EuroHPC-JU) under grant agreement No. 955776.
For this project, the EuroHPC-JU receives support from the European Union’s Horizon 2020 research and innovation programme of Spain under grants PCI2021-121934 (UPV) and PCI2021-121976 (UCLM), among other countries.

\bibliographystyle{./IEEEtran}
\bibliography{Biblio}

\end{document}